\documentclass[twocolumn]{aa}  

\usepackage{xcolor}
\usepackage{bm}
\usepackage{amsmath}
\usepackage{graphicx}

\usepackage{txfonts}
\usepackage{dsfont}
\newcommand{\esp}[1]{\mathbb{E}\left( {#1} \right)}
\newcommand{\var}[1]{\mathrm{Var}\left( {#1} \right)}

\begin{document}

   \title{Statistical properties and correlation length in star-forming molecular clouds: I. Formalism and application to observations}
 \titlerunning{Statistical properties and correlation length: I. Formalism and application}

   \author{E. Jaupart \inst{1} \and G. Chabrier\inst{1}$^,$ \inst{2} }
   \authorrunning{Jaupart, Chabrier}
   \institute{Ecole normale sup\'erieure de Lyon, CRAL, Universit\'e de Lyon, UMR CNRS 5574, F-69364 Lyon Cedex 07, France\\
              \email{etienne.jaupart@ens-lyon.fr}
         \and
             School of Physics, University of Exeter, Exeter, EX4 4QL, UK\\
             \email{chabrier@ens-lyon.fr} }

   \date{Received 14/04/2021; Accepted 12/05/2022}

 
  \abstract{Observations of molecular clouds (MCs) show that their properties exhibit large fluctuations. 
  The proper characterization of the general statistical behavior of these fluctuations, from a limited sample of observations or simulations, is of prime importance to understand the process of star formation.
In this article, we use the ergodic theory for any random field of fluctuations, as commonly used in statistical physics, to derive rigorous statistical results. We  outline how to evaluate the autocovariance function (ACF) and the characteristic correlation length of these fluctuations. We then apply this statistical approach to astrophysical systems characterized by a field of density fluctuations, notably star-forming clouds. When it is difficult to determine the correlation length from the empirical ACF, we show alternative ways to estimate the correlation length. Notably, we give a way to determine the correlation length of density fluctuations from the estimation of the variance of the volume and column-density fields. We show that the statistics of the column-density field is hampered by biases introduced by integration effects along the line of sight  and we explain how to reduce these biases. 
The statistics of the probability density function (PDF) ergodic estimator  also yields the derivation of the proper statistical error bars. We provide a method that can be used by observers and numerical simulation specialists to determine the latter. We show that they  (i) cannot be derived from simple Poisson statistics and (ii) become increasingly large for increasing density contrasts, severely hampering the accuracy of the high end part of the PDF because of a sample size that is too small.
 As templates of various stages of star formation in MCs, we then examine  the case of the Polaris and Orion B clouds in detail. We calculate, from the observations, the ACF and the correlation length in these clouds and show that the latter is on the order of $\sim$1\% of the size of the cloud. This justifies the assumption of statistical homogeneity when studying the PDF of star-forming clouds. These calculations provide a rigorous framework for the analysis of the global properties of star-forming clouds from limited statistical observations of their density and surface properties.}


   \maketitle
%

\section{Introduction}

Observations of molecular clouds (MCs) show that their main properties (velocity and column-density) exhibit large fluctuations.  These fluctuations are at the heart of the star formation process (\citealt{padoan2002,maclow2004,hennebelle2008,hopkins2012}), implying that knowledge of their statistical characteristics is of prime importance. The accurate determination of the statistics of any quantity 
must rely on either a large enough number of samples or a large enough sample, so that a natural question arises: can we derive accurate statistical properties of MCs from observations, 
and if so how can we evaluate the level of accuracy? The relevance of a general statistical analysis of the global properties of MCs (e.g., mass, density PDF, temperature and velocity dispersion) deduced from observations 
and numerical simulations for studies of star formation processes must be assessed properly. 
For example, all of the theories that aim at determining the mass spectrum, that is the  initial mass function (IMF) or the star formation rate (SFR) in a molecular cloud, rely on the assumption 
that a restricted number of observations or numerical simulations are representative of any MCs with similar properties  (see e.g., \citealt{hennebelle2008,hopkins2012,vazquez2019} and reference therein). This key assumption must be tested.

Indeed, in studies of star formation based on observations or numerical simulations, one only has access to a small number of samples (and in reality only one most of the time). Therefore, in order
to evaluate the statistics of the various stochastic fields of interest, one makes the basic assumption, sometimes called the "fair-sample hypothesis," that the available sample is large enough for volumetric (or time) averages 
to be meaningful (see e.g., \citealt{Peebles1973} for a discussion in the context of cosmology). This assumption is only valid for stochastic fields that are statistically homogeneous and ergodic \citep{papoulis1965}. 
Here, one should note that statistical homogeneity must not be confused with spatial homogeneity (we come back to this point below). 
The assumption of statistical homogeneity has been adopted by many authors, for example in studies of turbulent flows with or without self gravity \citep{chandra1951,chandra1951gravity,batchelor1953,pope1985,frisch1995,pan2018,Pan2019A,Pan2019B,jaupart2020,jaupart2021generalized} 
and in cosmology for studies of the dynamics of structures in the Universe \citep{Peebles1973,heinesen2020}. This assumption, however, provides no information on the magnitude of fluctuations around the average.

Quantifying (i) whether the "fair-sample hypothesis" is correct and, if so, (ii) what are the statistical error bars that derive from it, is an important issue when addressing PDF determinations in star-forming clouds. This is related to the completeness of the observations which we reformulate in this study in terms of statistical accuracy. In the context of the PDF of column-densities, \citet{alves2017}, for instance, made an attempt to illustrate this problem by introducing the concept of open and closed density contours. These authors suggest that complete observations, that is considered as statistically significant and worth studying, correspond to closed contours. Although interesting, such an approach, however, can not be considered as a robust statistical determination of the bias and the statistical errors corresponding to incomplete observations of a cloud PDF. It is one of the very aims of this article to provide such a robust statistical analysis, using standard tools of random field theory and signal processing. Notably, one of the goals of the paper is to identify the statistical properties of the density field in a cloud inferred from column density data, and to derive procedures, based on the aforementioned tools, to accurately estimate these properties, whatever the PDF (lognormal or not). As such our approach does not make any assumption about the initial driving mechanism of the random motions responsible for the PDF of a cloud: turbulence or gravity.

The proper and standard way of addressing the previous issue relies on Ergodic theory. Ergodic Theory allows one to circumvent the problem of dealing with a single sample and to derive a robust measure of the accuracy of field statistics derived from the available data. In the present context, it also enables us to assess and quantify the relevance of a statistical approach on the evolution of star forming MCs. 
The key quantity is the correlation length, which is defined in terms of the integral of the auto-covariance function (see e.g., \citealt{papoulis1965}). 
The fundamental result is that ergodic estimates are accurate if the dimensions of the sample, $i.e.$ a whole cloud or part of it, are large enough compared to the correlation length. 
A proper determination of the correlation length in MCs is therefore of prime importance.

In pioneer works, \citet{scalo1984, kleiner1985} studied the correlations of centroid velocities of the $\rho$-Oph and Taurus complex respectively and only found evidence of weak correlations at short scale on the order of their resolution.  \citet{kleiner1984} studied the correlation of the column density field of the Taurus complex in search of a statistically significant length scale characterizing the separation of condensations within the complex but did not perform an evaluation of the correlation length, as defined above.

The objectives of this article are twofold.  First, our main objective aims at examining the relevance and validity of a statistical approach based on ergodic theory to study the stochastic fields of star-forming MCs. 
Second, we seek to identify which statistical properties of the density field can be inferred from column density data and we derive procedures to obtain accurate estimates of these properties. The article is organized as follows.
In \S2, we outline the mathematical framework that yields the definition of the auto-covariance function (ACF) and correlation length of any statistical sample. In \S3 we derive ways to determine the correlation length of any stochastic field without having to compute the ACF. In \S4, we examine the case of astrophysical stochastic fields induced by compressible turbulent motions. In \S5, we focus on the case of star-forming clouds and on the ways to infer the statistics of these fields from observations of column-densities. In \S6, we apply our calculations to the typical star-forming cloud Polaris. We identify artifacts that are generated when one uses the statistical properties of the column-density field to infer those of the real density field; we show how to reduce these biases. In particular we derive a procedure to obtain proper error bars for the column density PDF. In \S7, we examine the case of Orion B. Section 8 is devoted to the conclusion.



\section{Methods: Mathematical framework for a statistical approach}
\label{sec:Math} 

As mentioned in the introduction, a statistical approach of the properties of a cloud (or part of)  is valid if this latter is large enough, compared to the correlation length of the quantity of interest, for the measured statistical quantities  to be representative with high confidence of the genuine quantities. How to measure this confidence level and thus the relevance of a statistical approach is given by the ergodic theory, as commonly used in statistical physics or in the study of dynamical systems. Indeed, ergodicity implies by definition that different observations and realizations of a given statistical quantity yield results comparable enough for each of them to be  representative of the average real quantity. It is described in the next section.

\subsection{Ergodic theory}

We rederive here some ergodic theorems that lead to the definition of the correlation length. Let us consider a (scalar) stochastic field $X(\bm{y})$, which depends on a $D$-dimensional position vector $\bm{y}$ ($D=1$, $2$ or $3$). For a specific and fixed $\bm{y}$,  $X(\bm{y})$ is a random variable of which  we want to accurately determine the statistics. 

\subsubsection{Frequency interpretation and repeated trials}

The usual way to estimate the statistical average or expectation $\esp{X(\bm{y})}$  of the random variable $X(\bm{y})$ is to observe $N$ samples $X(\bm{y}, \omega_i)$, $1\leq i \leq N$, of $X(\bm{y})$ and to build the unbiased estimator
\begin{equation}
    \hat{X}_{\bm{y},N} = \frac{1}{N} \sum_{i=1}^N X(\bm{y}, \omega_i), \label{eq:freqestimate}
\end{equation}
with variance
\begin{equation}
    \mathrm{Var}(\hat{X}_{\bm{y},N})= \sigma(\hat{X}_{\bm{y},N})^2 = \frac{\mathrm{Var}(X(\bm{y}))}{N} = \frac{\sigma(X(\bm{y}))^2}{N}, \label{eq:varfreqestimate}
\end{equation}
where $\sigma$ is the standard deviation (std). From Bienaymé-Tchebychev inequality \citep{papoulis1965}, we know that,  for any real number $m>0$,
\begin{equation}
   \mathbb{P}\left( |\hat{X}_{\bm{y},N} -  \esp{X(\bm{y})} | \leq m \, \sigma(\hat{X}_{\bm{y},N}) \right) \geq 1 - \frac{1}{m^2},  \label{eq:tchebyfreqestimate}
\end{equation}
where $\mathbb{P}$ denotes the probability of a given event. We note that this inequality is valid for any random field, whether it is Gaussian or not. Although Tchebychev inequality gives a lower limit for the probability, it allows to give a confidence interval  to measure the accuracy of the estimator given by Eq.~(\ref{eq:freqestimate}). The larger the number of samples $N$, the smaller the std $\sigma(\hat{X}_{\bm{y},N})$ and the more accurate the estimate in Eq.~(\ref{eq:freqestimate}). 

In case of statistical homogeneity for field $X$, the expectation $\esp{X(\bm{y})}$ and std $\sigma(X(\bm{y}))$ are no longer functions of the positions  and one can drop the reference to $\bm{y}$ in Eqs. (\ref{eq:varfreqestimate}) and (\ref{eq:tchebyfreqestimate}).

\subsubsection{Ergodic theorems,  autocovariance function, correlation length} \label{subsec:ergodictheorems}

In the context of the study of MCs, one usually has only a single sample of $X(\bm{y})$. As mentioned in the introduction, if one wants to be able to describe the stochastic fields at play, one assumes statistical homogeneity and build the  unbiased estimator
\begin{equation}
    \hat{X}_L = \frac{1}{L^D} \int_{\Omega} X(\bm{y}) \, \mathrm{d} \bm{y}, \label{eq:defestimergodic}
\end{equation}
where $\Omega=[-\frac{L}{2},\frac{L}{2}]^D$ is a control volume of linear size L and volume $L^D$, which we want to be as large as possible\footnote{The following calculations are made with this particular type of cubic control volume, as the calculations are easier to follow. We give in App.~(\ref{app:ergodiccalculationgeneralvolume}) the calculations for a control volume of any shape.}.  The ergodic estimator $\hat{X}_L$ has a variance
\begin{equation}
    \mathrm{Var}( \hat{X}_L ) = \frac{1}{(L)^D} \int_{[-L,L]^D} C_X(\bm{y}) \, \prod_{k=1}^D \left( 1 - \frac{|y_k|}{L} \right) \, \mathrm{d} \bm{y} \label{eq:varXl},
\end{equation}
where $C_X(\bm{y})= \esp{X(\bm{y}' + \bm{y}) \, X(\bm{y}')}-\esp{X}^2$ is the auto-covariance function (ACF) of $X$ at a lag $\bm{y}$. The stochastic field $X$ is said to be mean ergodic if the estimator $\hat{X}_L$ converges  toward  $\esp{X}$ as $L \rightarrow \infty$ either in the mean square (MS) sense, meaning:
\begin{equation}
\esp{|\hat{X}_L -\esp{X}|^2 } =  \mathrm{Var}( \hat{X}_L ) \xrightarrow[L \rightarrow \infty]{} 0,  
\end{equation}
or in probability meaning that,  for every $\epsilon >0$, 
\begin{equation}
  \mathbb{P}\left( |\hat{X}_{L} -  \esp{X)} | > \epsilon) \right) \xrightarrow[L \rightarrow \infty]{} 0.
\end{equation}
Bienaymé-Tchebychev inequality (Eq \ref{eq:tchebyfreqestimate}) not only shows that if $X$ is MS mean ergodic $\hat{X}_L$ also converges in probability, but it also provides a confidence interval for the estimate  $\hat{X}_L$. Slutsky's theorem allows to write an equivalence for the ergodicity of $X$ in a more convenient form: indeed,  $X$ is MS mean ergodic if and only if  
\begin{equation}
    \frac{1}{(L)^D} \int_{[-L,L]^D} C_X(\bm{y}) \mathrm{d} \bm{y} \xrightarrow[L \rightarrow \infty]{} 0.  
\end{equation}
From there we  obtain two sufficient (physical) conditions for $X$ to be mean ergodic. Either:
\begin{equation}
    \int_{\mathbb{R}^D} C_X(\bm{y}) \mathrm{d} \bm{y} < \infty,
\end{equation}
or
\begin{equation}
   C_X(\bm{y}) \xrightarrow[{| \bm{y}| \rightarrow \infty}]{}   0.
\end{equation}
We assume both, and use the common  definition of the correlation length $l_c(X)$ of the field $X$ as a function of the ACF \citep{papoulis1965}:
\begin{equation}
    (l_c(X))^D = \frac{1}{2^D\, C_X(\bm{0})}\int_{\mathbb{R}^D} C_X(\bm{y})  \mathrm{d} \bm{y}. \label{eq:defLc}
\end{equation}
This definition generalizes the usual definitions for $1$D fields:
\begin{equation}
    l_c(X) =  \frac{1}{C_X(\bm{0})}\int_{[0,+\infty[} C_X(y)  \mathrm{d} y = \frac{1}{2\,C_X(\bm{0})}\int_{\mathbb{R}} C_X(y)  \mathrm{d} y. 
\end{equation}
For $ l_c(X)\ll L$ we then have from Eq (\ref{eq:varXl}) 
\begin{equation}
    \mathrm{Var}( \hat{X}_L ) \simeq \mathrm{Var}(X) \left(\frac{ 2 \, l_c(X)}{L}\right)^D = \mathrm{Var}(X) \left(\frac{ l_c(X)}{R}\right)^D,
 \label{eq:varergodicestimate}
\end{equation}
where $R=L/2$. If we compare Eq.~(\ref{eq:varergodicestimate}) with Eq.~( \ref{eq:varfreqestimate}), we see that instead of having the number of samples, $N$, we now have  the ratio $(R/l_c)^D$,  where $R$ (or $L$) is 
usually an observationally accessible quantity. We can thus interpret the ratio $(R/l_c)^D$ as an effective number of "independent" samples.  This result is of prime importance in the analysis of fluctuations within any stochastic field.

\noindent Furthermore, the correlation length is linked to the value of the power spectrum $\mathcal{P}_X(\bm{k})$ of $X$ at $\bm{k}=\bm{0}$. Indeed
\begin{eqnarray}
    (l_c(X))^D &=&  \frac{1}{2^D \, C_X(\bm{0})} \int_{\mathbb{R}^D} C_X(\bm{y})  \mathrm{d} \bm{y}, \nonumber \\
    &=& \frac{1}{2^D \, C_X(\bm{0})} \, \mathcal{P}_X(\bm{0}).
\end{eqnarray}

\subsubsection{Ergodic hypothesis and ergodic theory}

The results of ergodic theory derived above enable us to define under which conditions volumetric averages correspond to statistical averages and to provide a confidence interval for the estimate  $\hat{X}_L$ of the expectation of $X$. 
However, these results rely on the knowledge of the statistical properties of $X$ and more precisely of its ACF, which is in general not known. 
To apply this theory to the study of a real field, such as the density field for example, one must use ergodicity as an assumption. The above results can then be used to test the validity of this assumption 
and the accuracy of the estimates that are derived from it in a self-consistent way.

\subsection{Estimates of the autocovariance function and correlation length } \label{subsec:ACFestim}

As shown in the previous section, the knowledge of the ACF of $X$ (or of the value of the power spectrum of $X$ at $\bm{k}= \bm{0}$) is of crucial importance to measure the relevance of a statistical approach 
in studies of the properties of  large (astrophysical) systems. In practice, however, the ACF of $X$ must be evaluated from data.

\subsubsection{Reliability of the estimators of the auto-covariance and the power spectrum.} \label{sec:reliabilityACF}

In most cases, data are drawn from a finite size sample so that the ACF is not reliable at large lag (large scales). To simplify the notation, we now introduce the variable $X_\mu = X - \esp{X}= X - \mu$ 
and define the estimate,  for a sample of size $L$,
\begin{eqnarray}
\hat{C}_{X}^L(\bm{y})  = \frac{1}{\prod_i \left(L-|y_i|\right)} \iiint_{-R+\frac{|y_i|}{2}}^{R-\frac{|y_i|}{2}}  X_\mu\left(\bm{u}-\frac{\bm{y}}{2}\right)\, X_\mu\left(\bm{u}+\frac{\bm{y}}{2}\right) \,\mathrm{d} \bm{u}&& \\
=  \frac{1}{\prod_i \left(L-|y_i|\right)} \iiint_{-L+|y_i|}^{L-|y_i|}  X_\mu\left(\frac{\bm{u}-\bm{y}}{2}\right)\, X_\mu\left(\frac{\bm{u}+\bm{y}}{2}\right) \,\frac{\mathrm{d} \bm{u}}{2^D}. && \label{eq:unbiasedACF}
\end{eqnarray}
This is an unbiased estimate of $C_X(\bm{y})$ but its variance is  increasing as $|y_i| \rightarrow L$ and eventually becomes very large due to poor sampling. We thus introduce the biased estimate
\begin{eqnarray}
\hat{C}_{X,L}(\bm{y}) = \frac{\prod_i \left(L-|y_i|\right)}{L^D} \, \hat{C}_{X}^L(\bm{y}), \label{eq:biasedACF}
\end{eqnarray}
which is still a good estimate at small scales compared to $L$ and has a reduced variance. We note however that it is an unbiased estimator of the quantity entering the integral in Eq.~(\ref{eq:varXl}). 
Finally, it is also the Fourier Transform of the periodogram $S_L$  which is defined as:
\begin{equation}
S_L(\bm{k}) = \frac{1}{L^D} \left| \int_{[-\frac{L}{2},\frac{L}{2}]^D} X(\bm{y}) e^{i \bm{k} \cdot \bm{y}} \mathrm{d} \bm{y}\right|^2.
\end{equation}
It is the usual estimate of the power spectrum of $X$, $\mathcal{P}_X$. It is, however, a biased estimator of the power spectrum $\mathcal{P}_X$ and is only  unbiased asymptotically, in the limit $L\rightarrow \infty$. 
Moreover, the variance of the estimator $S_L$ does not vanish as $L\rightarrow \infty$ \citep{papoulis1965}, which makes it quite unreliable. 

We thus see that, because of the finite size of the sample, one cannot obtain a reliable estimate of the ACF (or of the power spectrum) for all lag values. 
Furthermore, in many cases, the mean value of $X$ is not known and is replaced in Eq.~(\ref{eq:unbiasedACF}) by its estimate $\hat{X}_L=\hat{\mu}$, which introduces further, but reasonable, bias 
(see \citealt{papoulis1965} for a more complete discussion). 

\subsubsection{Periodic estimators}

To get rid of the effect of finite sampling, one may perform simulations in periodic calculation boxes or may artificially add some periodicity to the available data to obtain the following estimate:
\begin{eqnarray}
\hat{C}_{X,\mathrm{per}}(\bm{y})=\frac{1}{L^D} \int_{[-\frac{L}{2},\frac{L}{2}]^D} \left(X_{\hat{\mu}}(\bm{y} + \bm{u}) \, X_{\hat{\mu}}(\bm{u}) \right) \mathrm{d} \bm{u},
\end{eqnarray}
where $X_{\hat{\mu}} =X - \hat{\mu} = X -\hat{X}_L $ and where one makes the identification $X_{\hat{\mu}}(\bm{y} + \bm{n} L)= X_{\hat{\mu}}(\bm{y})$. However, in such cases, the spatial average of the estimated ACF is necessarily $0$. Indeed
\begin{eqnarray}
\int_{[-\frac{L}{2},\frac{L}{2}]^D} \frac{\hat{C}_{X,\mathrm{per}}(\bm{y})}{L^D} \mathrm{d} \, \bm{y} &=&  \int_{([-\frac{L}{2},\frac{L}{2}]^D)^2} \frac{\left(X_{\hat{\mu}}(\bm{y} + \bm{u}) \, X_{\hat{\mu}}(\bm{u}) \right)}{L^{2D}} \mathrm{d} \bm{u} \, \mathrm{d}\bm{y} \nonumber \\ 
&=& \left(\int_{[-\frac{L}{2},\frac{L}{2}]^D} \frac{X_{\hat{\mu}}(\bm{u})}{L^D} \mathrm{d} \bm{u} \right)^2 = 0,
\end{eqnarray}
due to the assumption that $X_{\hat{\mu}}$ is periodic. 

Therein lies a significant problem: as the correlation length is defined as an integral over all possible lags, it is not easy to evaluate the reliability of estimates that are obtained in this manner.  

Therefore, one  traditionally produces an estimate for $l_c$ (or the integral scale $l_i$)  in either of the next two ways. Either one searches for the $e^{-1}$ value of the reduced ACF $\hat{C}_X/\var{X}$ to obtain an estimate of the correlation length (see e.g. \citealt{kleiner1984,kleiner1985}), assuming some exponential envelop for the ACF. Or, if the ACF decays fast enough at scales larger than $l_c(X)$, as is the case in turbulence (see previous section), the ACF is then generally extrapolated with a decaying exponential in regions where it becomes non monotonic (see e.g. \citealt{batchelor1953,reinke2016,reinke2018}) so one can perform the integral and give a reliable estimate of $l_c(X)$ if $l_c(X) \ll L$.

 \section{Fluctuations and estimation of the correlation length} \label{sec:fluctuationsandcorr}

\subsection{Expected fluctuations in repeated trials} \label{subsec:repeatedtrials}

Be it for (numerical) experiments that can be repeated several times or for a statistically homogeneous and  stationary field,  volume averaged quantities fluctuate around their true expectations. In the former case, the  volume averaged quantities fluctuate between the different samples while in the latter case they fluctuate in time. As we show in the following, these fluctuations depend on the ratio  $(l_c/R)$. By studying these, we thus aim to obtain an accurate estimate of  $(l_c/R)$, without having to calculate the ACF.

We consider here the case where one can reproduce several times the same experiment, as can be done for instance with numerical experiments  or as can be approximated for clouds with similar conditions. We wish to determine the expected amplitudes of fluctuations of  volume averaged quantities between samples. 

To each experiment $i$ of the $N$ trials corresponds a value of the estimate $\hat{X}_{L,i}$ defined by Eq.~(\ref{eq:defestimergodic}). From Bienaymé-Tchebychev inequality, we know that $\hat{X}_{L,i}$ lies around the true expectation $\esp{X}$ within a distance such that, in probability,
\begin{equation}
    \mathbb{P}\left( |\hat{X}_{L,i} -  \esp{X} | \leq m \, \sigma(X)  \left(\frac{ l_c(X)}{R}\right)^{D/2} \right)\geq 1 - \frac{1}{m^2}.  \label{eq:tchebyergodicestimate}
\end{equation}
The average over the $N$ trials (or sample average)
\begin{equation}
    \hat{X}_{L}^{(N)}= \frac{1}{N} \sum_{i=1}^N \hat{X}_{L,i} \label{eq:sampleaverageergo}
\end{equation}
is obviously a better estimate of $\esp{X}$ as
\begin{equation}
    \mathbb{P}\left( |\hat{X}_{L}^{(N)} -  \esp{X} | \leq m \, \frac{\sigma(X)}{\sqrt{N}}  \left(\frac{ l_c(X)}{R}\right)^{D/2} \right)\geq 1 - \frac{1}{m^2}.  \label{eq:tchebyergodicsampleestimate}
\end{equation}
We then expect the $\hat{X}_{L,i}$ to fluctuate around the sample average $\hat{X}_{L}^{(N)}$  with variance
\begin{equation}
\var{\hat{X}_{L,i}-\hat{X}_{L}^{(N)}} =  \sigma(X)^2 \, \left(\frac{ l_c(X)}{ R}\right)^{D} \left(1 - \frac{1}{N} \right), \label{eq:varrepeat}
\end{equation}
such that
\begin{equation}
    \mathbb{P}\left( |\hat{X}_{L,i} - \hat{X}_{L}^{(N)} | \leq m \, \sigma(X)  \left(\frac{ l_c(X)}{ R}\right)^{D/2} \left(1 - \frac{1}{N} \right)^{1/2}\right)\geq 1 - \frac{1}{m^2}.  \label{eq:tchebysampletosamplecestimate}
\end{equation}
If $l_c(X)$ is known, eqs. (\ref{eq:tchebyergodicestimate}), (\ref{eq:varrepeat}) and (\ref{eq:tchebysampletosamplecestimate}) allow one to give statistical error bars. Conversely, if $l_c(X)$ is not known, these equations give information on the product $\sigma(X) (l_c(X)/R)^{D/2}$ by performing the same statistical experiment several times  and studying the dispersion of the $\hat{X}_{L,i}$ around  $\hat{X}_{L}^{(N)}$. 

 Indeed, the half length $l_{50\%}$ of the segment centered on $\hat{X}_{L}^{(N)}$  within which $50\%$ of the estimate $\hat{X}_{L,i}$ falls,  verifies\footnote{If the statistics are Gaussian, we obtain an equality where we have $0.67$ instead of $\sqrt{2}$ in Eq.~(\ref{eq:estimatelcmultitrials}).}
\begin{equation}
    \sqrt{2} \sigma(X)  \left(\frac{ l_c(X)}{ R}\right)^{D/2} \left(1 - \frac{1}{N} \right)^{1/2} \lesssim l_{50\%}. \label{eq:estimatelcmultitrials}
\end{equation}
This gives a quick and easy qualitative determination of $\sigma(X) (l_c(X)/R)^{D/2}$. More quantitatively, the empirical variance of the sample of the $N$ trials
\begin{equation}
\mathrm{Var}^{(N)} = \frac{1}{N-1} \sum_{i=1}^{N} \left( \hat{X}_{L,i}-\hat{X}_{L}^{(N)} \right)^2,
\end{equation}
is an unbiased estimator of
\begin{equation}
 \var{\hat{X}_{L,i}} =\sigma(X)^2 (l_c(X)/R)^{D}.
 \end{equation}
 Computing the variance $\mathrm{Var}^{(N)}$ thus yields an easy and rigorous method to determine $(l_c(X)/R)^{D}$  and the correct error bars for statistical experiments.

\subsection{Expected temporal fluctuations for a statistically homogeneous and stationary field.} \label{subsec:stationnary}

 We consider here the case of a  statistically homogeneous and stationary field (i.e., whose statistical properties are invariant under space and time translations). This can for example describe a steady turbulent flow  (e.g. as simulated in a periodic box) without gravity. 

As before, to each time $t$ corresponds a value of the estimate $\hat{X}_{L}(t)$ defined by Eq.~(\ref{eq:defestimergodic}) which lies around the true expectation $\esp{X}$ within a distance such that, in probability,
\begin{equation}
    \mathbb{P}\left( |\hat{X}_{L}(t) -  \esp{X} | \leq m \, \sigma(X)  \left(\frac{ l_c(X)}{R}\right)^{D/2} \right)\geq 1 - \frac{1}{m^2}. 
\end{equation}
The time average over a large timescale $T$
\begin{equation}
    \hat{X}_{L,T}= \frac{1}{T}\int_{t_0}^{ t_0+T} \hat{X}_{L}(t) \, \mathrm{d} t, \label{eq:timeaverageergo}
\end{equation}
where $t_0$ is a time at which the steady state is reached, is a better estimate of $\esp{X}$ as it has a variance
\begin{equation}
\var{ \hat{X}_{L,T}} \simeq \sigma(X)^2  \left(\frac{ l_c(X)}{ R}\right)^{D} \frac{\tau_c(X_L)}{2T},
\end{equation}
where $\tau_c(X_L) \ll T$ is the correlation time. Then, the signal  $\hat{X}_{L}(t)$ will fluctuate around $ \hat{X}_{L,T}$ with an empirical (temporal) variance
\begin{equation}
\mathrm{Var}_T= \frac{1}{T}\int_{t_0}^{ t_0+T} \left( \hat{X}_{L}(t) -  \hat{X}_{L,T} \right)^2 \mathrm{d}t \label{eq:VarT}
\end{equation}
which, providing that  $\tau_c(X_L) \ll T$, is an accurate estimate of 
\begin{equation}
 \var{\hat{X}_{L}(t)} =\sigma(X)^2 (l_c(X)/R)^{D}. \label{eq:VarXLt}
 \end{equation}
Hence, computing the variance $\mathrm{Var}_T$ yields an easy and robust estimate of $(l_c(X)/R)^{D}$  and the correct error bars for a statistically stationary field.

\subsection{Fluctuations of integrated fields over a column.}

The previous cases work for experiments that can be repeated or for statistically stationary fields. However, in some situations, the two conditions cannot be fulfilled either because it is impossible to reproduce the experiment a large number of times or because the fields are not stationary (for instance in the presence of gravity). 

In that case an estimate of ratio $(l_c/R)$  can be obtained if one has access to the integral of field $X$ over a column of fixed length $L=2R$:
\begin{equation}
\Sigma_X(\bm{r}) = \int_{[0,L]} X(\bm{r},z) \mathrm{d} z,
\end{equation}
where $\bm{r}$ is a vector of $D-1$ dimension (typically 2). The column must have a constant length to avoid creating spurious biases (see \S \ref{sec:columndens}).

As $\Sigma_X/L$  corresponds to averaging $X$ along one direction, we thus expect that its fluctuations will be reduced in comparison of those of $X$. The longer the length $L$ of the column, the smaller the fluctuations of $\Sigma_X/L$ are expected to be. More quantitatively, the variance of $\Sigma_X/L$ is
\begin{eqnarray}
\var{\frac{\Sigma_X}{L}} &=& \frac{\var{\Sigma_X}}{L^2} = \frac{\esp{\Sigma_X(\bm{r}) \, \Sigma_X(\bm{r})}}{L^2} \\
                                      &=& \frac{1}{L^2} \int_{[0,L]^2} C_X(\bm{0},z-z') \mathrm{d}z \, \mathrm{d}z' \\
                                      &=&  \frac{1}{L} \int_{[-L,L]} C_\rho(\bm{0},u)  \left(1-\frac{|u|}{L} \right) \mathrm{d}u. 
\end{eqnarray}
A similar equation for the centroid velocities was given in \citet{scalo1984}. Now, if the ACF of $X$ is isotropic at short lags and $l_c(X) \ll L$, one can make the approximation
\begin{eqnarray}
\var{\frac{\Sigma_X}{L}} &\simeq&   \frac{1}{L} \int_{[-L,L]} C_\rho(|u|)  \mathrm{d}u, \\
                                       &\simeq&  \var{X} \frac{2 l_i(X)}{L},
\end{eqnarray}
where $l_i(X)$ is the integral scale (Batchelor 1953) which, in most cases, verifies $l_i(X) \simeq l_c(X)$ (see \citealt{jaupart2021generalized}).   Thus a quick and easy estimate of ratio $l_c(X)/R$ is given by 
\begin{eqnarray}
\frac{\var{\Sigma_X}}{L^2 \var{X}} &\simeq&   \frac{l_c(X)}{R}. \label{eq:estimateLcColumn}
\end{eqnarray}
This method was applied to the density field ($\rho = X$) in \citet{jaupart2021generalized}. The above estimate (Eq. \ref{eq:estimateLcColumn}) was shown to produce the  trends predicted analytically and thus to be a good approximation of the actual ratio $l_c(\rho)/R$.
\section{Application to  astrophysical fields} \label{subsec:applicationtoastro} 

The general results derived in \S\ref{sec:Math} can be applied to many physical and astrophysical systems. 
They have been used extensively in cosmology but have somehow been overlooked in studies of star formation. 
Today, it is generally accepted that star formation is triggered  by density fluctuations generated by compressible turbulence injected at a large scale in MCs (see e.g. \citealt{mckee2007} and reference therein).
In this context, the density field $\rho$ (or the logarithmic density field $s=\mathrm{ln}(\rho/\esp{\rho})$) is of prime interest and its cumulative distribution function (CMF) 
and probability density function (PDF) must be determined accurately. 

Each of these statistical quantities is associated with a stochastic field $X$ to which the results of \S\ref{sec:Math} can be applied. 
For instance, the CMF $F_{\rho}(\rho_0)$ at $\rho_0$ is linked to the stochastic field $h_{\rho_0}(\bm{y})= \Theta\left(\rho_0 - \rho(\bm{y}) \right)$ (where $\Theta$ is the Heavyside function), 
as
\begin{equation}
F_{\rho}(\rho_0) = \esp{h_{\rho_0}(\bm{y})}=\esp{h_{\rho_0}},
\end{equation} 
while the PDF $f_{\rho}(\rho_0)$ is given by $f_{\rho}(\rho_0) = \esp{\delta_{\rho_0}(\bm{y})}=\esp{\delta_{\rho_0}}$,
where $\delta_{\rho_0}(\bm{y})= \delta\left(\rho_0 - \rho(\bm{y}) \right)$ is Dirac's distribution. Usually the PDF is rather deduced from histograms with some bin size $\Delta \rho$ such that
\begin{equation}
f_{\rho}(\rho_0) \Delta \rho \simeq F_{\rho}(\rho_0 + \Delta \rho) - F_{\rho}(\rho_0) = \esp{h_{\rho_0 + \Delta \rho} - h_{\rho_0}}. 
\end{equation}

In principle, knowledge of the ACF of  all  these fields is required to establish the accuracy of the estimations. 
Fortunately, it can be shown that sometimes, with a few simplifying assumptions, one can proceed with the ACF of $\rho$, only, in some situations. 
This is explained in detail in App.~(\ref{app:ergodicCMFPDF}).


\subsection{Exact results regarding the properties of the auto-covariance function (ACF) of $\rho$}
For a  statistically homogeneous field,  the ACF of $\rho$, the density field, can be expressed in term of the second order structure function :
\begin{equation}
    S_{\rho}^{(2)}(\bm{y}) = \esp{ \left\lbrace\rho(\bm{u} + \bm{y}) - \rho(\bm{u}) \right\rbrace^2 },
\end{equation}
as $S_{\rho}^{(2)}(\bm{y}) = 2 \left( C_\rho (\bm{0}) - C_\rho(\bm{y} \right)$. 
A similar statement can  be made for the logarithmic density field $s=\mathrm{ln}(\rho/\esp{\rho})$.  This helps us to grasp some key features of the ACF. 
At very short scale (below the viscous scale), the density field is supposed to be differentiable and hence $C_\rho$ must possess second-derivatives at $\bm{y}=\bm{0}$. 
Then, due to the parity of the ACF and because it is maximal at $\bm{y}=\bm{0}$, its gradient must exist and be equal to $\bm{0}$ at $\bm{y}=\bm{0}$. 

Furthermore, \citet{jaupart2021generalized}, generalizing the work of \citet{chandra1951}, show that for a statistically homogeneous density field, the quantity 
\begin{equation}
\esp{\rho} \, \var{e^s} \, l_c(\rho)^3 = \frac{\var{\rho}}{\esp{\rho}}\, l_c(\rho)^3
\end{equation}
is an invariant of the dynamics.

\subsection{Phenomenology of (compressible) turbulence}

The phenomenology of compressible turbulence \citep{kritsuk2007} can be derived, with some adjustments, from that of incompressible turbulence \citep{frisch1995}. 
Thus, we use the latter to derive some expected features of the density ACF in star-forming MCs that can be described by such phenomenology.  

In isotropic turbulence, the second order structure function is observed to be a monotonic increasing function of separation distance, at least in the inertial range, 
and to converge rapidly toward   $2 \var{\rho}$ at scales that are larger than the integral scale $l_\mathrm{i}$. 
This integral scale  (not to be confused with the injection scale of turbulence, see \S \ref{subsec:injdiffcorr}) is defined in the same manner as the correlation length (\citealt{batchelor1953})
\begin{equation}
    l_\mathrm{i} = \frac{1}{C(\bm{0})} \int_0^\infty C(r) \mathrm{d}r .
\end{equation}
In many situations, $l_c \sim l_\mathrm{i}$, as shown in \citet{jaupart2021generalized}.
Thus, at small scales (short lags) and in the inertial range, the ACF must be a monotonically decreasing function. 
Above the inertial range, it is often assumed that the structure function and the ACF are still monotonic and the ACF is usually approximated by a decaying exponential, 
even though density fluctuations are likely to generate oscillations of the observed and estimated ACF as it tends to zero (\citealt{batchelor1953,reinke2016,reinke2018}). 

In compressible isothermal and stationary turbulence, the density field $\rho$ is found to be approximately lognormal \citep{kritsuk2007,federrath2010}, 
implying that the logarithmic density field $s=\mathrm{ln}(\rho / \esp{\rho})$ is Gaussian with variance $\sigma(s)^2=\mathrm{ln}(1+(b\mathcal{M})^2)$. 
In such Gaussian conditions, the ACFs of $\rho$ and $s$ are linked by the following equation:
\begin{equation}
    C_\rho(\bm{y}) = \esp{\rho}^2 \left( e^{C_s(\bm{y})} -1 \right).
\end{equation}
As a consequence, if $C_\rho$ (or $C_s$) is monotonically decaying toward   $0$, we deduce that:
\begin{equation}
    \left(\frac{\sigma(s)^2}{e^{\sigma(s)^2}-1}\right)^{1/3} l_c(s) \leq l_c(\rho) \leq l_c(s) \, ,
\end{equation}
where we have used the following inequalities: $e^{ax} -1 \leq x(e^{a}-1)$ for $0\leq x\leq 1$ and $ax \leq e^{ax} -1$ $\forall x$.
For typical star forming conditions, $\sigma(s)^2 \lesssim 4$, implying that:
\begin{equation}
    0.4 \, l_c(s) \lesssim l_c(\rho) \leq l_c(s),
\end{equation}
or,
\begin{equation}
    l_c(s) \sim l_c(\rho).
\end{equation}
This shows that under Gaussian conditions, for the two lengths $l_c(s)$ and $l_c(\rho)$, knowledge of one of them is sufficient to characterize the other one within an order of magnitude.

\subsection{Large injection scale but small correlation length} \label{subsec:injdiffcorr}

Density and velocity fluctuations in MCs are thought to originate from turbulent motions driven at large scale \citep{mckee2007,brunt2009turbulent}, i.e. at scale comparable to the cloud scale $L$. This means that the energy of these turbulent motions is injected at an injection scale $l_{\rm inj} \sim L$, below which the turbulent cascade eventually occurs. 

The injection scale $l_{\rm inj}$, however, is not the correlation length $l_c$ of either the velocity, kinetic energy or density fields. Otherwise, if $l_{\rm inj}= l_c$, every estimate produced from volumetric averages of the former fields would be inaccurate and far from the actual statistical values. This could result in large fluctuations of these averaged quantities either between different simulations (samples) or at different times for steady turbulent flows (see  \S \ref{sec:fluctuationsandcorr}). What is instead observed in numerical simulations of compressible and steady turbulent flow is that volume averaged quantities of these fields display rather small fluctuations around their mean values. This is the case, for example, for the  rms Mach number $\mathcal{M} = v_{\rm rms}/ c_s$, where $v_{\rm rms}$ is the root of the volume average square velocity $\bm{v}^2$ and $c_s$ the sound speed. To be more explicit,  
\begin{equation}
\mathcal{M}^2 =  \frac{1}{L^3} \int_{[-\frac{L}{2},\frac{L}{2}]^3} \frac{\bm{v}^2(\bm{y})}{c_s^2} \mathrm{d} \bm{y},
\end{equation}
 so the result of  \S \ref{sec:fluctuationsandcorr} can be applied with  $X = \bm{v}^2/c_s^2$ and $\mathcal{M}^2 = \hat{X}_L$.

 In \citet{federrath2013universality}, a series of numerical simulations of isothermal compressible turbulence driven to $\mathcal{M} \simeq 17$  with an injection scale $l_{\rm inj} = L/2 = R$ is presented. Once a statistical steady state is reached, the volume averaged Mach number $\mathcal{M}$ (or $\mathcal{M}^2$ which is a measure of the volume averaged specific kinetic energy) displays fluctuations that are rather small compared to their average values (their Fig. 1). Would we have $l_c(\bm{v}^2) = l_{\rm inj} =  R$, fluctuations of the signal $\mathcal{M}^2(t)$ would have yielded a temporal variance (Eq. \ref{eq:VarT}):
\begin{equation}
\mathrm{Var}_T \simeq \sigma^2\left(\bm{v}^2\right) \left( \frac{l_c(\bm{v^2})}{R}\right)^3 = \sigma^2\left(\bm{v}^2\right),
\end{equation}
from Eq. \ref{eq:VarXLt}. Since the statistics of $\bm{v}$ are close to being Gaussian (their Fig. A1), this would imply 
\begin{equation}
\mathrm{Var}_T \simeq 2 \mathcal{M}_T^4,
\end{equation}
where $\mathcal{M}_T$ is the average of the signal $\mathcal{M}(t)$ over a time $T$. This would thus yield large fluctuations incompatible with their Fig. 1. The actual temporal variance $\mathrm{Var}_T$ of signal $\mathcal{M}(t)^2$ in \citet{federrath2013universality} rather yields a ratio
\begin{equation}
\frac{l_c(\bm{v}^2)}{R} = \frac{l_c(\bm{v}^2)}{l_{\rm inj}} \simeq 0.1, 
\end{equation}
which shows that $l_c(\bm{v}^2) \ll l_{\rm inj} = R$.

Furthermore, \citet{jaupart2021generalized} used the estimate produced by Eq. \ref{eq:estimateLcColumn} to compute the correlation length of the density field $\rho$. They found that within a factor of order unity, 
\begin{equation}
l_c(\rho) \simeq \lambda_s = L / \mathcal{M}^2 \ll L = 2 \,  l_{\rm inj},
\end{equation}
where $\lambda_s$ is the sonic length which is found to be close to the average width of filamentary structures in isothermal turbulence \citep{federrath2016}.

The above results show that  a large injection scale does not imply a large correlation length and that, on the contrary, correlation lengths in star-forming MCs are small compared with the injection scale (see above and Jaupart \& Chabrier (2021).

\subsection{Practical assumptions regarding the ACF}

From the above results, we thus assume that the ACF decays rapidly at scales larger than the correlation length $l_c$ ($l_c\sim l_\mathrm{i}$,  the integral scale) and then that the defining integral Eq.~(\ref{eq:defLc}) can be calculated only up to a few $l_c$. Moreover,  
we assume that the ACF can be bounded by a decaying exponential $\exp(-|\bm{y}|/\lambda)$, where $\lambda \sim l_c$ above and in the inertial range to allow the computation of the correlation length (we note that such an exponential behavior is prohibited at very small scales due to the differentiability of $\rho$).

\section{Star-forming clouds. Column densities as tracers of the underlying density field} \label{sec:columndens}

\begin{figure*}[!t]
    \centering
    \includegraphics[trim=0 500 0 130, clip,width=\textwidth]{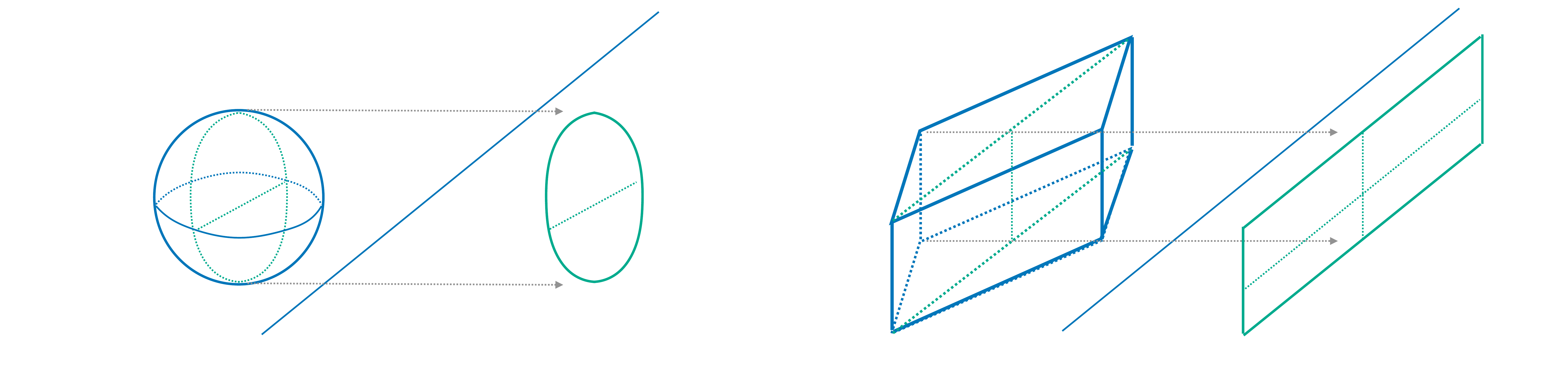}
    \caption{Projection of the two idealized situation Left panel: Case of a sphere. Right panel: Case of a cuboid mis-aligned with the line of sight. }
    \label{fig:projectionLOS}
\end{figure*} 

We now turn to observations of star-forming molecular clouds. Measurements provide values of the column-density $\Sigma(x,y)$, which is the integral of density along the line of sight ($\mathrm{l.o.s.}(x,y)$):
\begin{eqnarray}
    \Sigma(x,y) &=& \int_{\mathrm{l.o.s.}(x,y)} \rho(x,y,z) \, \mathrm{d}z \nonumber \\
    &=& \esp{ \rho } l(x,y) + \int_{\mathrm{l.o.s.}(x,y)} \delta \rho(x,y,z) \, \mathrm{d}z, \label{eq:calculSigmalos}
\end{eqnarray}
where $l(x,y)$ is the thickness of the cloud along the l.o.s. at $(x,y)$ and $\delta \rho = \rho - \esp{\rho}$ is the density fluctuation.  
Column densities are the only data that depend directly on the density field and one must determine how to retrieve reliable information from them.  

\subsection{Inhomogeneity and anisotropy due to integration over the line of sight }

Star forming clouds are shaped by turbulent motions conferring statistical properties to their geometrical characteristics, and hence to the area projected in a plane perpendicular to the line of sight 
and to the thickness projected along the line of sight. 
This is responsible for difficulties in evaluating exactly the statistical average of $\Sigma(x,y)$. However, provided that the cloud thickness is much larger than the correlation length, that is if $l(x,y) \gg l_c(\rho)$, 
we can reasonably assume that (see Eq.~(\ref{eq:calculSigmalos})):
\begin{equation}
\esp{\Sigma(x,y)} \simeq \esp{\rho} \esp{l(x,y)} .
\end{equation}
 One must note here that we are dealing with the statistical average and not with the spatial average. 
This equation shows that $\Sigma(x,y)$ may not be statistically homogeneous even if the density field $\rho$ is, just because of integration effects. 
To illustrate this important point, let us imagine two idealized situations. In one of them, the cloud is a sphere of radius $R$. In the other one, the cloud is a "cube" of side $L$ misaligned with the line of sight 
and seen from one of its edges such that the projected surface is of size $\sqrt{2} L \times L$ (see Fig.~(\ref{fig:projectionLOS})). For the sphere, the thickness along the line of sight is:
\begin{equation}
   \esp{ l_{\rm S(R)}}(x,y)= 2 R \left(1-\frac{x^2 + y^2}{R^2} \right)^{1/2}, \,\,\, x^2 + y^2 < R^2,
\end{equation}
whereas for the cubic cloud it is : 
\begin{eqnarray}
\esp{ l_{\rm C(L)}}(x,y) = \sqrt{2}L \left( 1 - \frac{\sqrt{2}|x|}{L}\right), \,\,\, |x|\leq \frac{L}{\sqrt{2}}, \, |y| \leq \frac{L}{2}.
\end{eqnarray}
Even though they are very simple, these two examples demonstrate that the column-density field may exhibit large-scale gradients 
and hence may not be statistically homogeneous, even if the density field is. Furthermore, as seen with the example of the cube, integration effects can also generate some anisotropy in the column-density field.

To reduce these effects, one can use a low pass filter to filter out large-scale gradients \citep{kleiner1984} and then treat the column density field as if it were homogeneous. Furthermore, most of the integration effects are expected to be produced by the first term of the r.h.s of Eq.~(\ref{eq:calculSigmalos}). Thus they are expected to affect column densities that are around  the (surface) average of the column density map $\left<  \Sigma \right>$. In contrast, high column density ($\Sigma(x,y) > \left<  \Sigma \right>$, the regions of interest for star formation) are most likely to originate from the second term of the r.h.s of Eq.~(\ref{eq:calculSigmalos}) and be produced by dense pockets along the line of sight. They are thus expected to be less affected by the integration effects and by the low pass filter. Studying the statistics of these high column density is thus expected to bear insights on the bias introduced by integration effects.  

In Sec. \ref{sec:Obs}, we will apply the  above considerations to the observations of the Polaris cloud. 

\subsection{Column-density field in a simulation box} \label{subsec:variancecolumndensity}

For a cubic simulation domain of size $L$, projecting the density field along one of the 3 principal directions of the cube leads to a statistically homogeneous column density field such that : 
\begin{equation}
 \esp{\Sigma(x,y)} = \esp{\rho} \times L. 
 \end{equation}
The results of \S \ref{sec:Math} and  \S \ref{sec:fluctuationsandcorr} can thus be applied with $X=\Sigma$ and the  ACF of $\Sigma$ in this case is
\begin{eqnarray}
    C_\Sigma(\bm{r}) &=& \esp{\left(\Sigma(\bm{u} + \bm{r})- \esp{\rho} \, L\right) \left(\Sigma(\bm{u})- \esp{\rho} \, L\right) } \nonumber\\
    &=& \int_{[-L/2,L/2]^2} C_\rho(\bm{r}, z-z')\, \mathrm{d} z \, \mathrm{d}z' \nonumber \\
    &=& \int_{[-L,L]} C_\rho(\bm{r},u) \mathrm{d}u \int_{-L+|u|}^{L-|u|} \frac{\mathrm{d} v}{2} \nonumber \\
    &=& L \int_{[-L,L]} C_\rho(\bm{r},u) \left(1-\frac{|u|}{L} \right) \mathrm{d}u.
\end{eqnarray}
Its variance is 
\begin{eqnarray}
\mathrm{Var}(\Sigma)&=&C_\Sigma(\bm{0}) = L \int_{[-L,L]} C_\rho(\bm{0},u)  \left(1-\frac{|u|}{L} \right) \mathrm{d}u.
\end{eqnarray}
Then, if the density field is statistically isotropic at small scales (i.e. the ACF is isotropic at short lag) and $l_c(\rho)\ll L$,
\begin{eqnarray}
\mathrm{Var}(\Sigma) &\simeq& 2\,L\, l_c(\rho)\, \mathrm{Var}(\rho). \label{eq:variancesigma}
\end{eqnarray}
 Then Eq.~(\ref{eq:variancesigma}) yields:
\begin{equation}
    \mathrm{Var}\left(\frac{\Sigma}{\esp{ \Sigma }}\right) \simeq \mathrm{Var}\left(\frac{\rho}{\esp{ \rho }}\right) \frac{2\,l_c(\rho)}{L} = \mathrm{Var}\left(\frac{\rho}{\esp{ \rho }}\right) \frac{l_c(\rho)}{R}, \label{eq:relationratiovar}
\end{equation}
which is a reformulation of Eq.~\ref{eq:estimateLcColumn}. This is an important result because it gives a  measure of $l_c(\rho)/R$ without having to compute the ACF.  \citet{brunt2010,federrath2010} for example found a ratio $\var{\Sigma/\esp{\Sigma}}/\var{\rho/\esp{\rho }}$ between $0.03$ and $0.15$. 

\citet{vazquez2001} were the first to study the impact of the $l_c(\rho)/R$ ratio on the statistics of column-density fields. Based on a crude interpretation of the central limit theorem (CLT), 
they proposed that, for $l_c(\rho)/R \rightarrow 0$, the  column-density PDF should appear to be Gaussian instead of lognormal. 
This is not consistent with the apparent lognormality of the observed column-density PDFs, which led these authors to conclude that $l_c(\rho)/R$ cannot be vanishingly small and that it must be on the order of $10^{-1}$. 
However, the CLT only applies to independent variables and can hardly be valid for the sum of correlated variables, even if correlations decay. This casts doubt on the conclusions of \citet{vazquez2001}. 
More recently, \citet{szyszkowicz2009} and \citet{beaulieu2011} have shown that, for some special types of correlations, the sum of a large number $N$ of lognormal variables tends to a lognormal distribution as $N \rightarrow \infty$. 
We conclude that knowledge of the $l_c(\rho)/R$ value does not allow robust conclusions on the shape of the column-density PDF. However, as shown by Eq.~(\ref{eq:relationratiovar}), 
the variance $\mathrm{Var}\left(\frac{\Sigma}{\esp{ \Sigma }}\right)$ does become vanishingly small as $l_c(\rho)/R$ tends to zero. In that case, one can show with high probability that:
\begin{equation}
    \mathrm{ln}\left(\frac{\Sigma}{\esp{ \Sigma }}\right) \simeq \frac{\Sigma-\esp{\Sigma}}{\esp{ \Sigma }}.
\end{equation}
Thus, in the limit of vanishing values of $l_c(\rho)/R$, the distributions of $\Sigma/\esp{\Sigma}$ and its logarithm are both Gaussian if one of them is.

\subsection{Decay  length of correlations}

We now examine how the decay of correlations of $\rho$ impacts the decay of correlations of $\Sigma$. 
For  sake of simplicity, we again consider the case of a cubic box in order to avoid unncessary complications. For the 2D field $\Sigma$, the correlation length is given by :
\begin{eqnarray}
l_c(\Sigma)^2 &= &\frac{1}{4} \frac{1}{\var{\Sigma}} \iint C_\Sigma(\bm{r}) \,  \mathrm{d} \bm{r} , \nonumber \\
                       &=& \frac{1}{4} \frac{1}{\var{\Sigma}}  \iint  L \int_{[-L,L]} C_\rho(\bm{r},u) \left(1-\frac{|u|}{L} \right) \mathrm{d}u  \,  \mathrm{d} \bm{r} , \nonumber\\
                       &\simeq& 2 \frac{L \var{\rho}}{\var{\Sigma}} l_c(\rho)^3. \nonumber\\
\end{eqnarray}
Using Eq.~(\ref{eq:variancesigma}), this implies that:
\begin{equation}
  l_c(\Sigma)^2 \simeq  l_c(\rho)^2 .
\end{equation}
This shows that correlations of the column-density fields are decaying over a characteristic length close to $l_c(\rho)$, the correlation length of the underlying density field. 
In general, we can thus assume that $l_c(\Sigma) \sim l_c(\rho)$, so that information gathered from the column-density yields an estimate of the characteristic decay length of correlations of the underlying density field $\rho$.  


\section{Application to the observations of the Polaris cloud} \label{sec:Obs}

As mentioned in \S\ref{sec:columndens}, observations trace back the column-density \citep{kleiner1984,schneider2015,ossenkopf2016}. These observations of the column-density PDFs in MCs show that regions where star formation has not occurred yet exhibit lognormal PDFs while regions with numerous prestellar cores develop power-law tails (PLTs) at high column densities \citep{kainulainen2009,schneider2013}. In addition to the integration effects yielding potentially the observed column-density to be anisotropic and inhomogeneous, observational data suffer further biases due to line of sight (l.o.s) contamination and noise \citep{schneider2015,ossenkopf2016}. L.o.s contamination causes two important biases. The observed power-law tail appears to be steeper than its corrected and uncontaminated counterpart while the observed variance  in the lognormal part appears to be smaller than its corrected counterpart (\cite{schneider2013}). The overall effect of l.o.s contamination is to produce an underestimation of the total variance of the column-density.

\begin{figure}[!t]
    \centering
    \includegraphics[width=\columnwidth]{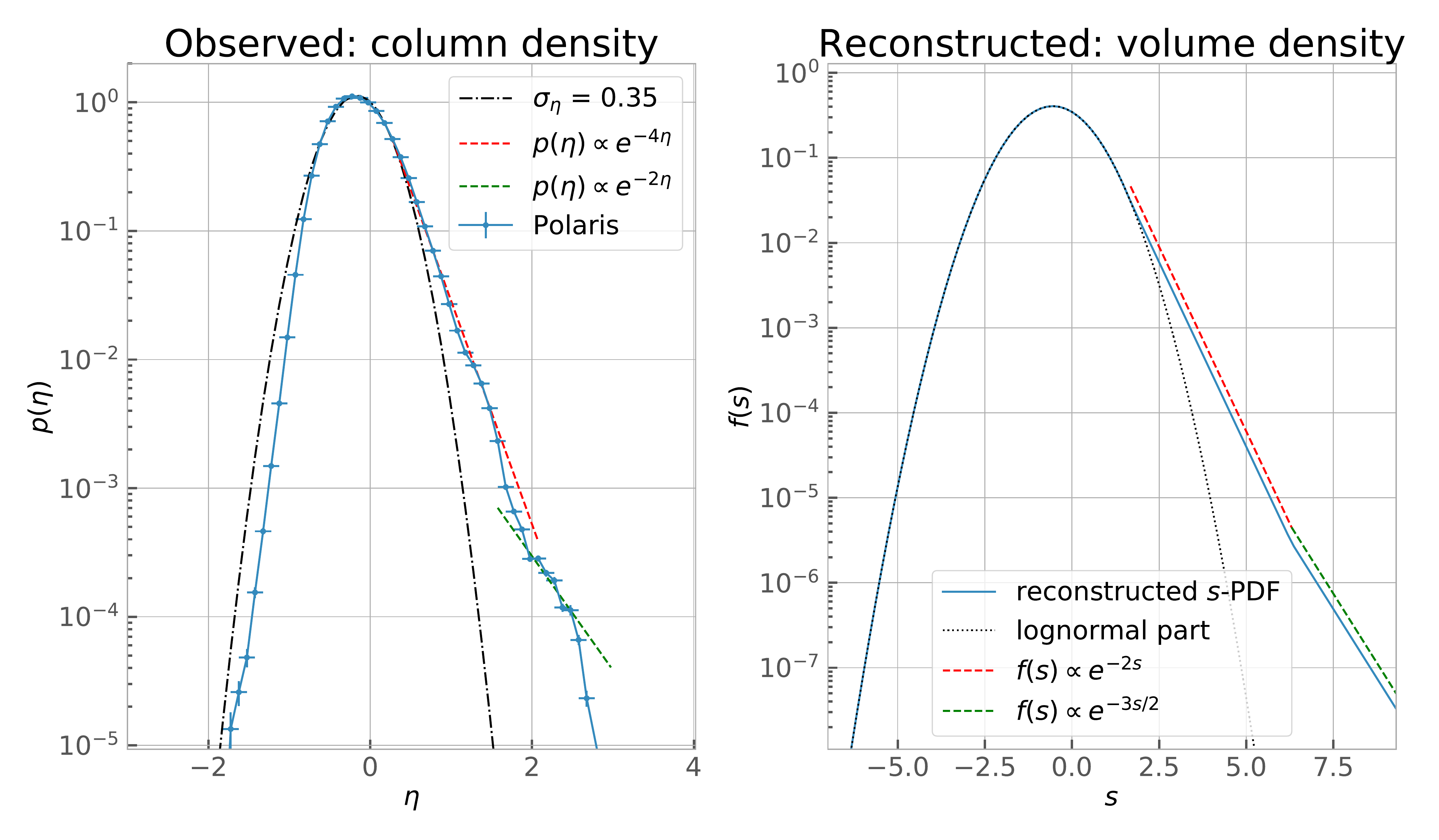}
    \caption{Column and volume density PDF of the Polaris cloud. Left: Observed logarithmic column-density ($\eta = \mathrm{ln}(\Sigma/\left< \Sigma \right>$)-PDF \citep{schneider2013,jaupart2020}. Right: Estimated and reconstructed underlying logarithmic density ($s=\mathrm{ln}(\rho/\esp{\rho})$)-PDF with the procedure from \cite{jaupart2020}.}
    \label{fig:PolarisPDF}
\end{figure}

\subsection{Polaris}

\begin{figure*}
    \centering
    \includegraphics[trim=0 150 0 150, clip,width=\textwidth]{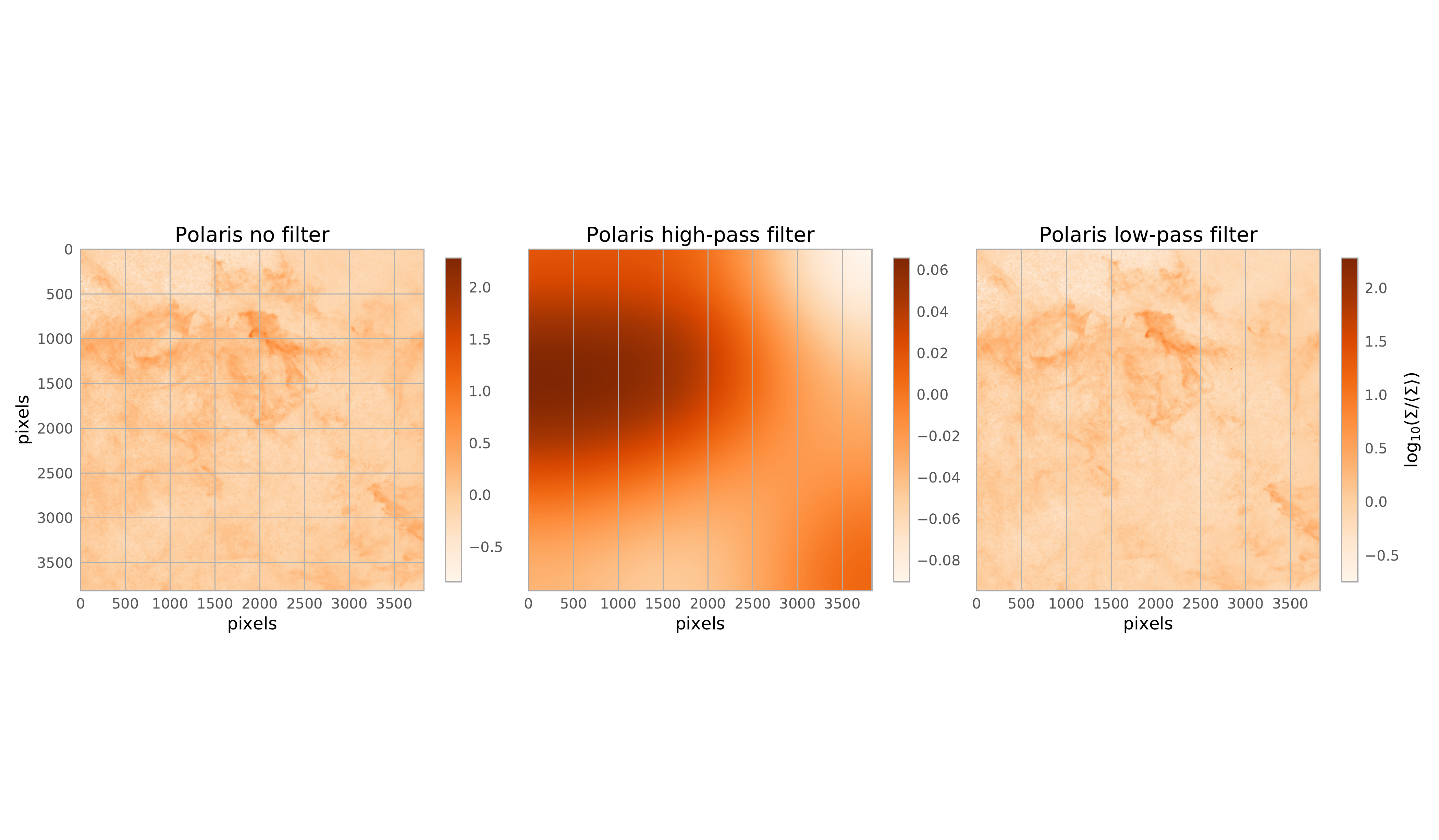}
    \caption{Column-density maps of the Polaris cloud. Left panel: without filter. Middle panel: with a high-pass filter filtering scales smaller than $L/2$. Right panel: with a low pass filter filtering scales larger than $L/2$. The low pass filter does not alter qualitatively the richness of structures found in the Polaris flare, while the high-pass filter shows a large-scale gradient that can be produced by an integration effect.}
    \label{fig:Polaris3filtimage}
\end{figure*}
\begin{figure*}
    \centering
    \includegraphics[trim=0 150 0 150, clip,width=\textwidth]{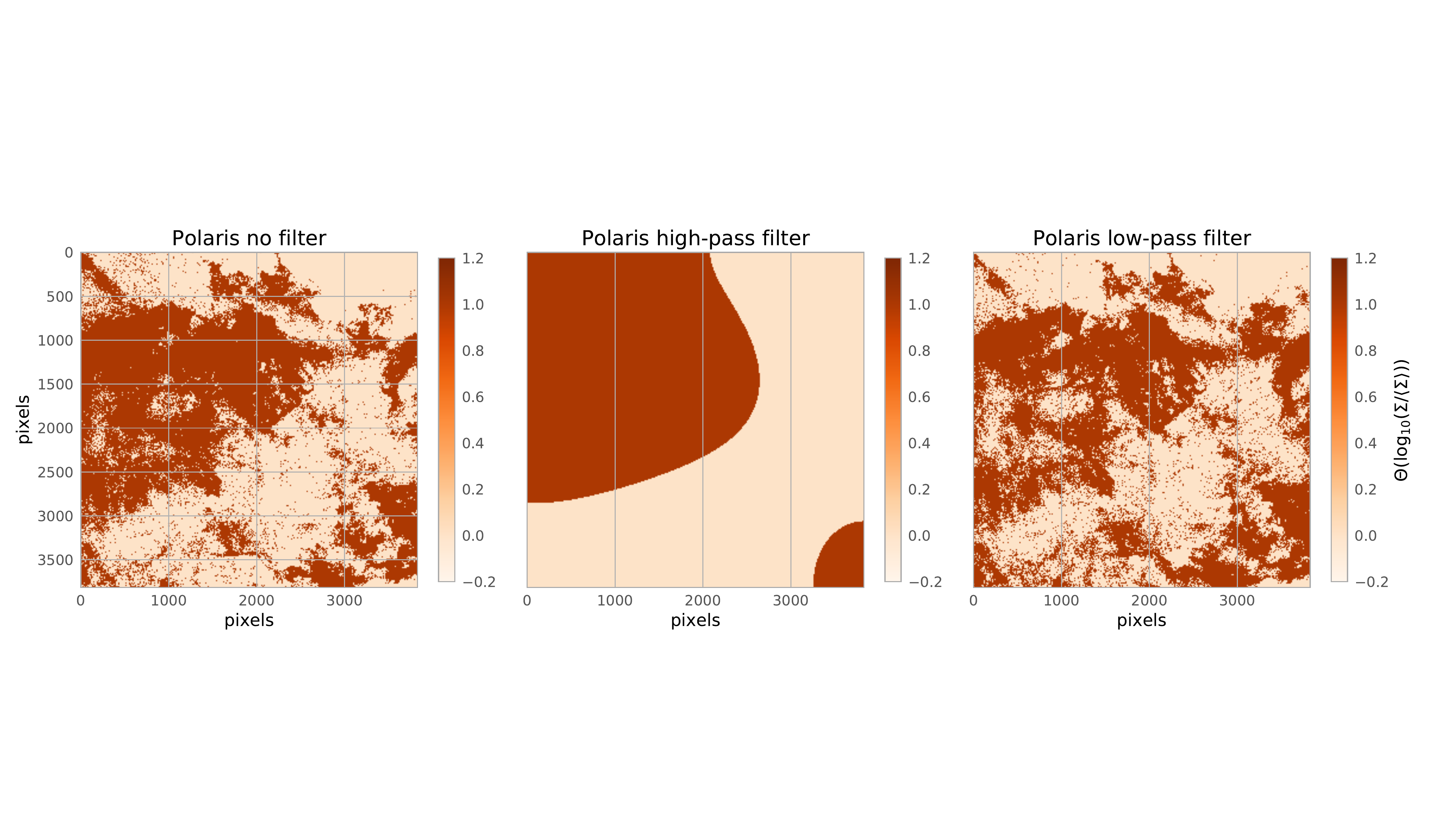}
    \caption{Same as Fig.~(\ref{fig:Polaris3filtimage}), but for the binary map $\Theta(\mathrm{log}(\Sigma/\left< \Sigma\right>))$ where $\Theta$ is Heaviside's step function. Regions where $\Sigma > \left< \Sigma \right>$ appear darker than regions where $\Sigma < \left< \Sigma \right>$.}
    \label{fig:PolarisHeaviside}
\end{figure*}

As a typical example of initial conditions of star formation in MCs, we focus on the Polaris flare, where line of sight contamination appears to be negligible \citep{andre2010,miville2010,schneider2013}. 
Furthermore, most of the stellar cores in this cloud are still unbound \citep{andre2010}, showing that star formation activity is very recent. 
Polaris is therefore a good candidate to probe the statistics of  initial phases of star formation in MCs. 

Data from Herschel Gould Belt survey extend across part of this cloud over approximately a 10 square degrees region with a linear size $L\sim 10 $ parsecs (pc) \citep{andre2010}. 
The cloud total mass and area above an extinction $A_v \geq 1$ are $M_{c, A_v \geq1}=1.21 \times 10^3 M_\odot$ and $A_{c, A_v \geq1} = 3.9\, \mathrm{pc}^2$, respectively. 
Dust temperatures are in a narrow $T_{\rm dust}= 13 \pm 1$K interval, indicating fairly isothermal conditions with an average Mach-number $\mathcal{M} \simeq 3$ \citep{schneider2013}.

The Polaris logarithmic column-density field $\eta$, where $\eta =\mathrm{ln}(\Sigma/\left< \Sigma \right>)$,  has  a PDF with an extended Gaussian part and two emerging power-law tails, 
a first one with exponent $\alpha_{\eta,1} \simeq -4$ followed by a shallower one with exponent $\alpha_{\eta,2} \simeq -2$ (Fig.~\ref{fig:PolarisPDF}). 
\citet{jaupart2020}  have shown that the first steep PLT is due to gravity beginning to affect turbulence in parts of the cloud and hence records an early stage of (local) collapse. Furthermore the authors outlined a procedure to reconstruct the underlying logarithmic volume density PDF, noted $s$-PDF, where $s =\mathrm{ln}(\rho/\esp{\rho})$, from data on the $\eta$-PDF. The underlying $s$-PDF displays a Gaussian part and two PLTs with exponents $\alpha_{s,1}=-2$ and $\alpha_{s,2}=-3/2$, respectively (see Fig.~\ref{fig:PolarisPDF}).

\subsection{Filtering large-scale gradients} \label{sec:filter}

As seen in \S\ref{sec:columndens}, integration effects can produce large-scale gradients and break statistical homogeneity as well as isotropy in the column density field. \\
Filtered and unfiltered column-density maps of the Polaris flare  are displayed in Figs.~\ref{fig:Polaris3filtimage} and \ref{fig:PolarisHeaviside}. 
The low pass filter does not alter qualitatively the intricate structures that exist, while the high-pass filter reveals a large-scale gradient likely due to integration effects. 
In order to partially reduce measurement artifacts, we use a low pass filter that screens out structures larger than $L/2$ in the column-density contrast ($\Sigma-\left<\Sigma \right>$), where $L$  is the  linear size of the observed region and we recall that $\left<  \Sigma \right>$ is the (surface) average of the column density map. 
We can then treat the column-density field as if it was homogeneous.

\begin{figure}
    \centering
    \includegraphics[width=\columnwidth]{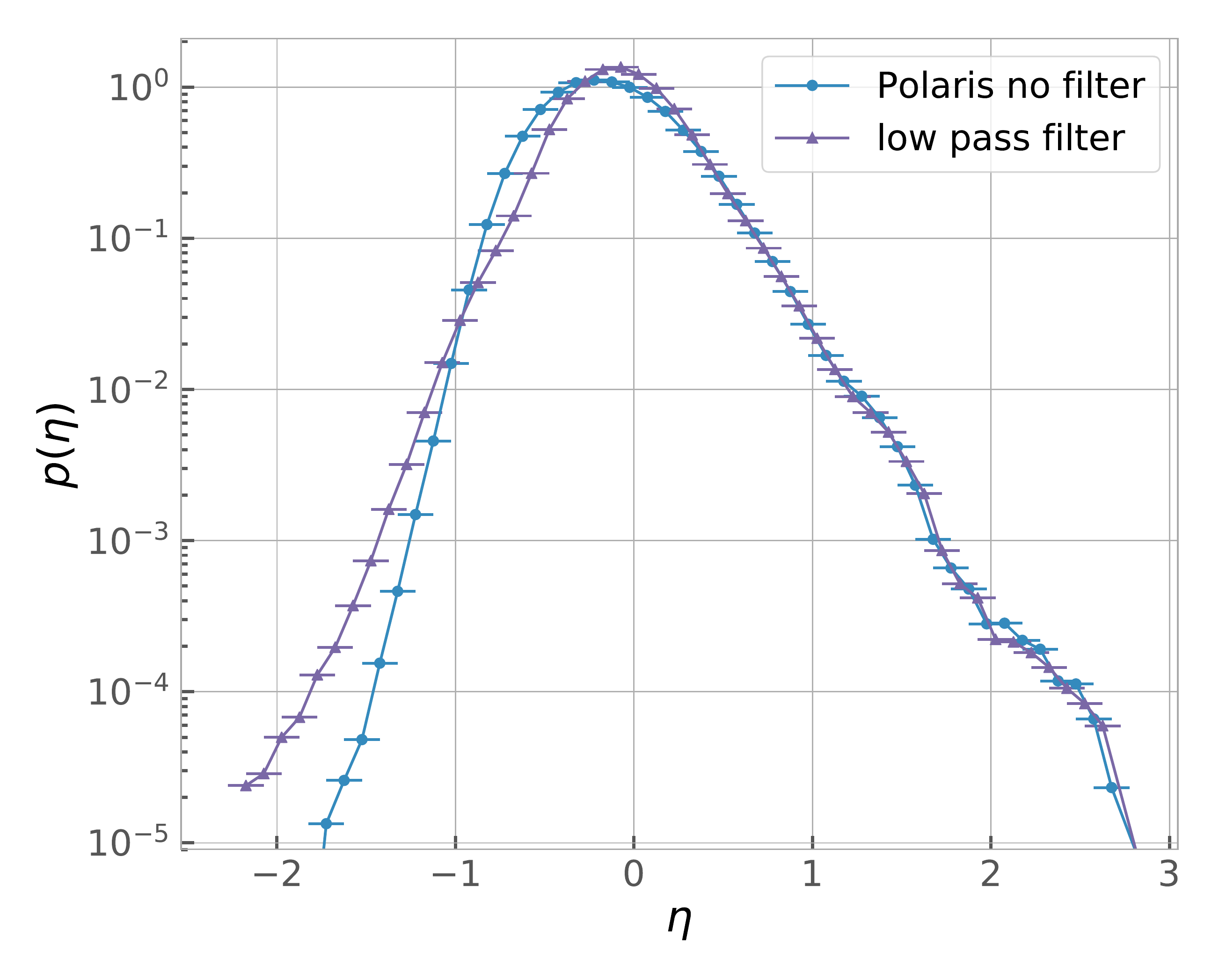}
    \caption{$\eta$-PDFs. Blue round and purple triangular symbols represent the PDFs of the unfiltered and low pass filtered maps, respectively. The filter does not alter regions with $\eta>0$ but increases the occurrence of regions with $\eta <-1$. Horizontal errorbars represent bin spacing.}
    \label{fig:pdffilteredpolaris}
\end{figure}

The low pass filter slightly diminishes the variance $\var{\Sigma/\left<\Sigma\right>}$ which is $\simeq 0.20$ and $\simeq 0.17$ for the unfiltered and low pass filtered data, respectively. 
It barely affects structures with a positive column-density contrasts but increases the occurrence of highly negative column-density contrasts. 
This is seen in Fig.~\ref{fig:pdffilteredpolaris}, which portrays the $\eta$-PDFs of the unfiltered and low pass filtered column-density maps. As mentioned in  \S\ref{sec:columndens}, high column density are not expected to be sensitive to integration effects.

\begin{figure*}
    \centering
    \includegraphics[trim=0 150 0 150, clip,width=\textwidth]{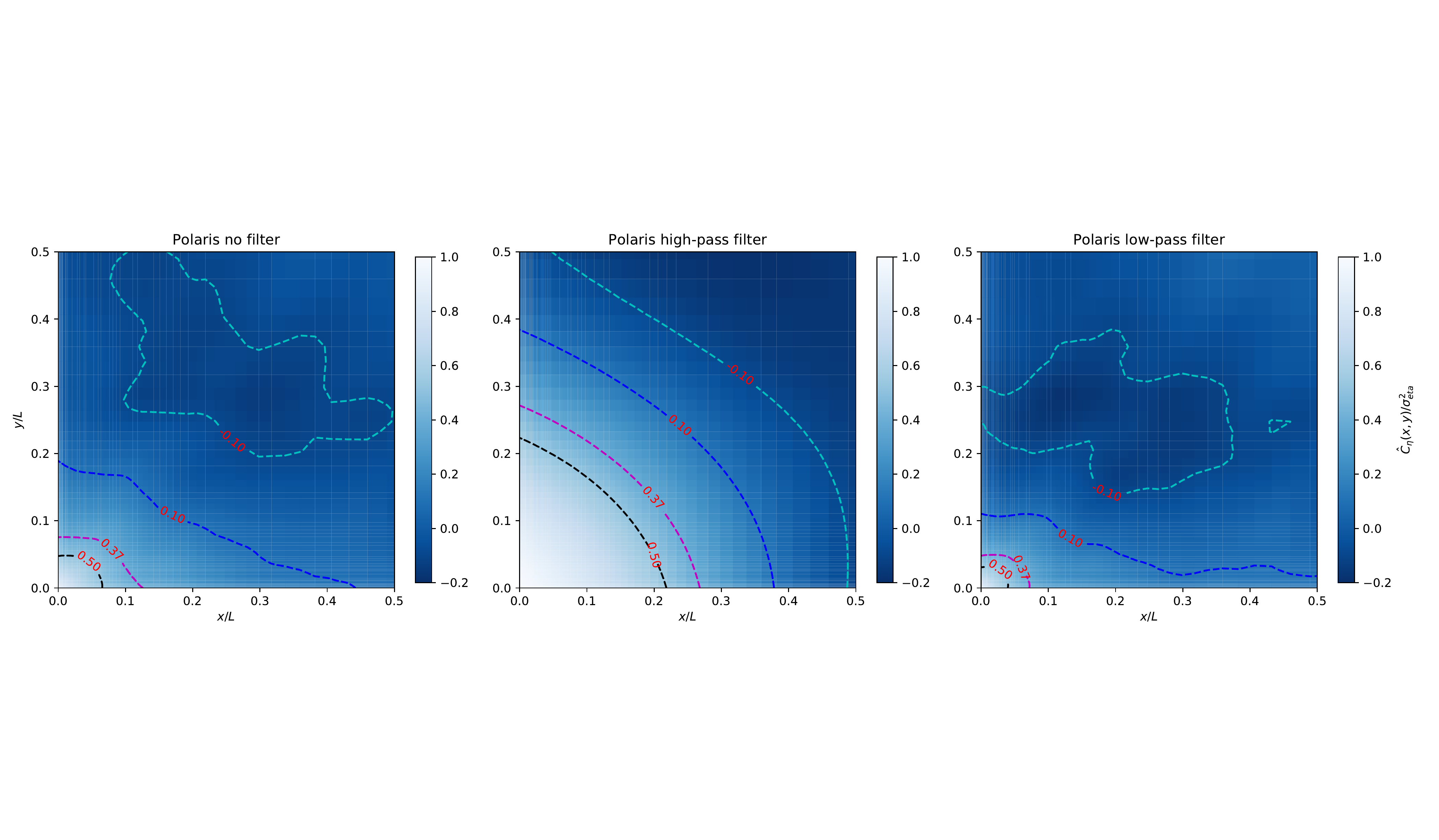}
    \caption{Reduced ACF function of $\eta$ ($\hat{C}_\eta / \var{\eta}$) for the Polaris flare. Left panel: without filter. Middle panel: with a high pass filter filtering scales smaller than $L/2$. Right panel: with a low pass filter filtering scales larger than $L/2$. Contours from black to purple to blue to light blue give the value of the reduced ACF at $0.5$, $e^{-1}\simeq 0.37$, $0.1$, $-0.1$.}
    \label{fig:Polaris3ACF}
\end{figure*}

\subsection{Estimated ACF and correlation length} \label{subsec:ObservedACF}

\begin{figure}[!t]
    \centering
    \includegraphics[width=\columnwidth]{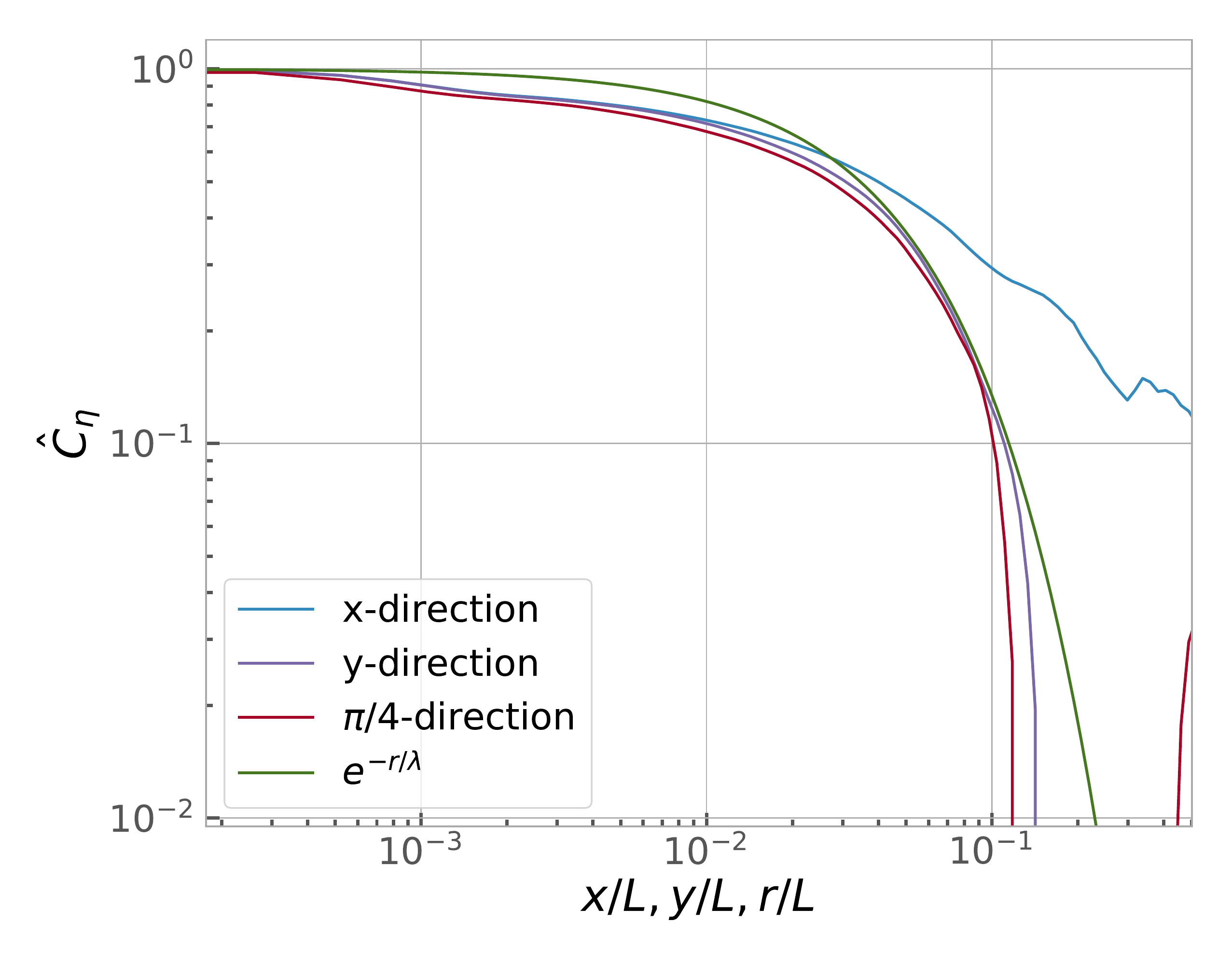}
    \caption{Reduced ACF of the low pass filtered map in three different directions. Blue line: $x$-direction ($y=0$). Purple line: $y$-direction $(x=0)$. Red line: $\pi/4$ or $x=y$-direction. Green line: exponential profile giving an estimate of the rate of decay (here $\lambda/L \simeq 5  10^{-2} $). A strong anisotropy is present in the $x$ direction at large scales ($x/L \geq 2 \, 10^{-2}$).}
    \label{fig:ACF3dir}
\end{figure}
\subsubsection{Correlation length of $\eta$ from the ACF}

We now estimate the ACFs of the logarithmic column-density field $\eta= \mathrm{ln}(\Sigma/\left<\Sigma\right>)$ for the three data sets (unfiltered, low and high pass filtered), using Eq.~(\ref{eq:biasedACF}). 
The 2D heat-maps of the reduced ACFs $\hat{C}_\eta / \var{\eta}$  are given in Fig.~\ref{fig:Polaris3ACF}. We only display the top-right quadrant of possible lags $(x>0,y>0)$ which amounts to half of the space useful to study the ACF due to its symmetry. The high pass filtered ACF illustrates the bias that can be introduced by integration effects. 

The three ACFs  all seem to be fairly isotropic at very short lags (scales)  but  are anisotropic at large ones.  The low pass filtered ACF seems to decay more rapidly at short lags with a reduce anisotropy than the unfiltered one. 
Fig.~\ref{fig:ACF3dir} displays the reduced ACF of the low pass filtered map in three different directions, $x$ ($\theta=0$), $x=y$ ($\theta=\pi/4$) and $y$ ($\theta=\pi/2$). 
As can be seen from the heat maps but also from Fig.~\ref{fig:ACF3dir}, a strong anisotropy is detected  at large scales in the $x$ direction ($x/L \geq 2 \, 10^{-2}$), while the ACF if fairly isotropic at shorter lags. 
From the $y$-direction to the $\pi/4$-direction, the data seem to be fairly isotropic and bounded by an exponential with $\lambda/L \simeq 5 \times 10^{-2}$. 
Anisotropy is most pronounced along the $x$-direction and the resulting estimated correlation length $\hat{l}_c(\eta)$ is :
\begin{equation}
\hat{l}_c(\eta) \simeq 6 \times 10^{-2}\, L \simeq \frac{1}{2} (2 \pi)^{1/2} \lambda, \label{eq:estimatelcPol}
\end{equation}
or $\hat{l}_c(\eta)/R \simeq 1.2 \times 10^{-1}$, thus $\hat{l}_c(\eta)/R \sim  10^{-1}$.   We then use $\hat{l}_c(\eta)$ as an estimate of $l_c(\rho)$ to within an order of magnitude, such as 
\begin{equation}
l_c(\rho) \sim10^{-1} \, R.
\end{equation}
In fact, we expect Eq.~(\ref{eq:estimatelcPol}) to provide upper bounds for ratios $l_c(\eta)/R$ and $l_c(\rho)/R$, because integration artifacts are only partially cancelled by the low pass filter. 

\subsubsection{Correlation length of $\eta$ from Eq. (\ref{eq:estimateLcColumn})}

 To obtain an estimate of $l_c(\eta)$ we could also apply the results of Sec. \ref{sec:fluctuationsandcorr}  to the low pass filtered map. By integrating the column density map along the $x$ ($\theta=0$) or $y$  ($\theta=\pi/2$) direction and computing the variance of the resulting integrated field we can obtain with  Eq. \ref{eq:estimateLcColumn} two estimates  $\hat{l}_{c,x}(\eta)$ and $\hat{l}_{c,y}(\eta)$ of $l_c(\eta)$ within a factor of order unity. However, the estimated ACF displays a strong anisotropy in the $x$ direction at large scales ($x/L \geq 2 \, 10^{-2}$), so we expect the estimates to give rather different results. 
Computing the estimates yields
\begin{eqnarray}
\hat{l}_{c,x}(\eta) \simeq 2.6 \, 10^{-1} R \\
\hat{l}_{c,y}(\eta) \simeq 3.5 \, 10^{-2} R, \label{eq:estimatelcY}
\end{eqnarray}
with $\hat{l}_{c,x}(\eta)> \hat{l}_{c,y}(\eta)$, as can be expected from the anisotropy of the observed ACF. Since the anisotropy of the ACF starts before it is significantly smaller than the variance, it is not clear which of the two estimates gives the better approximation of the  actual $l_c(\eta)$. However, they are both within a factor $3$ of the estimate produced by Eq.~(\ref{eq:estimatelcPol}) from the ACF which yielded $\hat{l}_c(\eta) \sim  10^{-1} R$. 

Using  $l_c(\eta)$ as an estimate of $l_c(\rho)$  yields again, to within an order of magnitude, $l_c(\rho) \lesssim10^{-1} \, R.$

\subsubsection{Correlation length of $\rho$ from Eq. (\ref{eq:estimateLcColumn}) or (\ref{eq:relationratiovar}) and the variance of $\Sigma$}

As discussed in Sec.~(\ref{subsec:variancecolumndensity}) and Eq.~(\ref{eq:relationratiovar}), one can also estimate the ratio $l_c(\rho)/R$ by (1) computing the variance $\var{\Sigma/\esp{\Sigma}}$, 
(2) giving an estimate of $\var{\rho/\esp{\rho}}$ and (3) giving an estimate of the average thickness of the cloud (the length of the line of sight), for example by assuming that the cloud has roughly the same dimension in the three directions.

\begin{figure*}[!t]
    \centering
    \includegraphics[trim=35 20 20 35, clip,width=\textwidth]{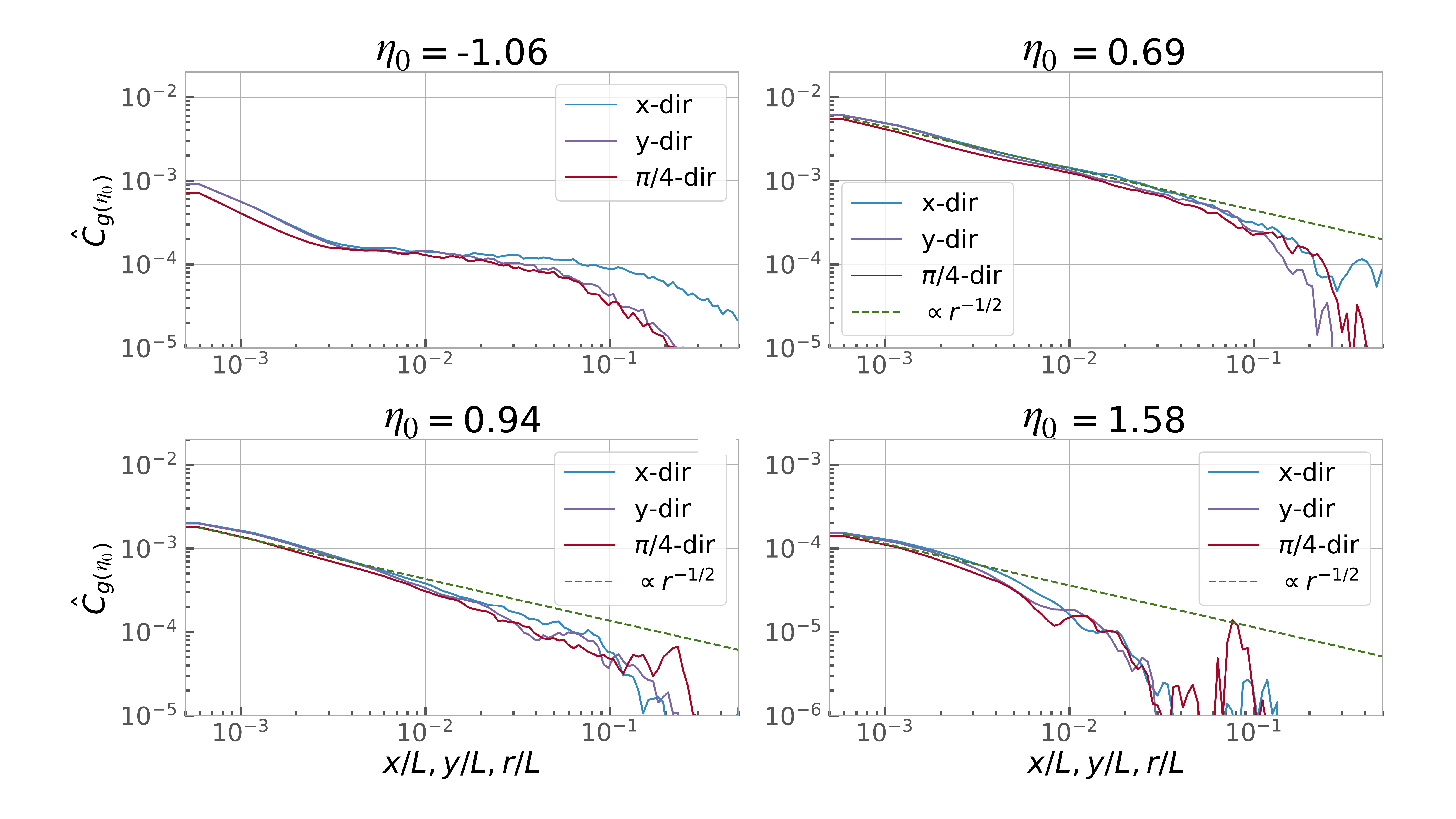}
    \caption{Estimated ACF of the field $g_{\eta_0}(\bm{y})$ for different values of $\eta_0=\eta$ in 3 different directions. Blue, purple and red lines represent respectively the $x$, $y$ and $\pi/4$ ($x=y$) directions. 
    The two top panels are for $\eta_0=-1.06$ and $0.69$, whereas the two bottom panels are for $\eta_0=0.94$ and $1.58$.  At low column-densities ($\eta_0=-1.06)$,  
    a strong anisotropy is detected in the $x$-direction and becomes noticeable at $x/L \geq 2 \, 10^{-2}$ as was the case for the column-density ACF  (see Fig.~(\ref{fig:ACF3dir})). 
    For high column-densities ($\eta_0>0$), however, the anisotropy is subdued and the ACFs are fairly isotropic at small scales up to $x/L, \, r/L \sim  \, 10^{-1}$ where the data become quite noisy. 
    Green dashed lines show the profile of an isotropic ACF proportional to $r^{-1/2}$ that  matches the data at short scales fairly well, at least over a decade.}
    \label{fig:ACFofPDFestimators}
\end{figure*}

In pure isothermal turbulence, $\var{\rho/\esp{\rho}} \simeq (b\mathcal{M})^2$, which is $\simeq 1$ for the Polaris case ($b \simeq 0.3$--$0.4$, $\mathcal{M} \simeq 3$). However, when gravity starts generating power-law tails in the density PDF, 
the variance becomes larger than $(b\mathcal{M})^2$ \citep{jaupart2020}. For Polaris, the column-density PDF displays a power-law tail with exponent $\alpha_\eta \simeq -4$, 
which is linked to an underlying density PDF with a power law tail exponent $\alpha_s \simeq -2$ (\citealt{federrath2013,jaupart2020}). 
Using the reconstructed $s$-PDF of Fig.~(\ref{fig:PolarisPDF}) from the procedure described in \citet{jaupart2020}, we can derive an estimate of $\var{\rho/\esp{\rho}}$. 
In principle, for such a model PDF, the variance is infinitely large due to the power-law tails exponents $\alpha_{s,1} = -2$ and $\alpha_{s,2}=-3/2$. 
However, we expect a cut-off at high (column)-density, which is indeed visible in the data. This cutoff may be due to a change of thermodynamic conditions of the cloud, e.g. from isothermal to adiabatic conditions. 
For a typical  cut-off number-density $n_{\rm ad} = 10^{10}$ $\mathrm{cm}^{-3}$ \citep{masunaga2000,machida2006,vaytet2013,vaytet2018} and for a cloud of average density $\overline{n} =10^3$ $\mathrm{cm}^{-3}$, 
the cutoff occurs at $s_{\rm ad} \simeq 16$. However, there may be other causes for a high density cut-off. 
In order to assess this possibility, we thus determine three different estimates of the variance $\var{\rho/\esp{\rho}}$ from the reconstructed $s$-PDF of Fig.~\ref{fig:PolarisPDF}: one densities up to 6.3 ($s \leq 6.3$), 
which corresponds to the onset of the 2nd PLT,  a second one for $s\leq 8$ in order to include contributions from the 2nd PLT, and a third one for $s \leq 16 \simeq s_{\rm ad}$ in order to include all the data up to the adiabatic limit. 
We obtain $\var{\rho/\esp{\rho}} \simeq 5$, $7$, $227$, respectively, such that: 
\begin{equation}
l_c(\rho)/R \simeq 0.04, \,\, 0.03, \,\, 0.001. \label{eq:estimationLCPolvar}
\end{equation}
This provides us with the conservative estimate $l_c(\rho)/R \sim 10^{-2}$, which is an order of magnitude smaller than the value estimated from the ACF (Eq.~(\ref{eq:estimatelcPol}))  but closer to the estimate Eq. (\ref{eq:estimatelcY}).

It is thus important to understand whether most of the anisotropy in the ACF originates from some integration artifacts and whether it causes or not an overestimation of the correlation lengths of $\eta$ or $\rho$.


\subsection{Ergodic estimate of the observed PDF, real error bars, and reduced integration artifacts}

As mentioned earlier, column-density PDFs serve as tracers of the statistics of the underlying density field. 
The various forms of these PDFs can be attributed to the various processes that are operating in MCs, from a fully lognormal distribution when purely turbulent motions dominate 
to a lognormal distribution with high density PLTs when gravitational effects become significant \citep{vazquez1994,passot1998,kainulainen2009,schneider2013}. This calls for a precise determination of the statistical uncertainty on the observed PDF, especially at high-density values.

The empirical PDF $\hat{f}_X(\xi_0)$ of stochastic field $X$ (here $X$ will be the column density $\eta$) is deduced from histograms with some bin size $\Delta \xi$. Error bars are usually estimated from Poisson statistics (using the number of points per bin) and can therefore be very small \citep{schneider2013}. It is worth delving deeper into this issue. A histogram yields the following estimate:
\begin{equation}
 \hat{f}_{X;L}(\xi_0) \Delta \xi \simeq \hat{F}_{X;L}(\xi_0 + \Delta \xi) - \hat{F}_{X;L}(\xi_0), 
\end{equation}
where $\hat{F}_{X;L}$ is the empirical cumulative distribution function. Formally, this amounts to the ergodic estimate of the average of the following field, noted $g_{\xi_0}(\bm{y})$:
\begin{eqnarray}
   g_{\xi_0}(\bm{y}) &=& h_{\xi_0 + \Delta \xi}(X(\bm{y})) - h_{\xi_0}(X(\bm{y})), \\
   \mathrm{where} \, \, h_{\xi_0}(X(\bm{y})) &=& \Theta(\xi_0 - X(\bm{y})),
\end{eqnarray}
(see Sec.~(\ref{subsec:ergodictheorems}) and (\ref{subsec:applicationtoastro}))). 
Thus, proper statistical error levels must be calculated using the results of Sec.~(\ref{sec:Math}) and in general  are not given by Poisson statistics. 

In App.~\ref{app:PDF}, we study in detail ergodic estimates of average quantities. In general, the correlation length of $g_{\xi_0}$ is a function of $\xi_0$ itself. 
For Gaussian (or lognormal) distributions, an important result is that the confidence interval becomes quite large for values $|\xi_0-\esp{X}|\geq \sigma(X)$, resulting in large errors if the sample size is too small. Thus, a reliable evaluation of the statistics of rare events (away from the average) requires very large sample sizes.

\subsubsection{Reduced integration effects at high density contrasts} \label{subsec:reducedeffects}

In this study, we focus on the column-density field $X =\eta$ and its PDF, noted $p(\eta)$. 
Using $g_{\eta_0}(\bm{y})$ and its ACF for various values $\eta_0$, we are able to determine the appropriate statistical error bars and to get rid of some of the artifacts that are due to integration along the line of sight. 
In practice, we expect that such artifacts are not significant in high column-density regions (see \S \ref{sec:filter}).  
For example, anisotropy of the Polaris column density ACF in the $x$-direction is likely due to integration effects (see Sec.~\ref{subsec:ObservedACF}). 

We expect, however, that the ACF of field $g_{\eta_0}$ for $\eta_0 >0$ is expected to show a reduced anisotropy at short scales. 
We thus obtain an empirical ACF of $g_{\eta_0}$ using Eq.~(\ref{eq:biasedACF}).  Fig.~\ref{fig:ACFofPDFestimators} displays the estimated PDF of $g_{\eta_0}$ for the low pass filtered column-density map. 
At low column-density ($\eta_0=-1.06$),  a strong anisotropy is observed in the $x$-direction starting at $x/L \geq 2 \times 10^{-2}$, as for the ACF of $\eta$ (see Fig.~\ref{fig:ACF3dir}). 
For positive column density contrasts ($\eta_0>0$), this anisotropy is reduced and the ACFs are fairly isotropic at small scales in both the $x$ and $\theta=\pi/4$ directions, up to $x/L \, \sim  \, 10^{-1}$ 
and $r/L  \,\sim  \, 10^{-1}$, respectively, where $r$ denotes separation distance in the $\theta=\pi/4$ direction. At larger separation distances, the data become quite noisy. This is consistent with the fact that the low path filtering procedure does not modify the PDF significantly in regions where $\eta >0$ (see Fig ~\ref{fig:pdffilteredpolaris}).

This suggests that most of the $\eta$ ACF anisotropy in the $x$-direction at scales in the $10^{-2} - 10^{-1}$ range is due to integration effects. The peak of the correlation in the $\pi/4$-direction at high column-densities ($\eta_0 = 1.58$) is probably due to the presence of the "Saxophone"-shaped filamentary structure that may be seen at the top of Fig.~\ref{fig:Polaris3filtimage}, which hosts most of the Polaris high density regions \citep{schneider2013}.

\subsubsection{Statistical errorbars} \label{subsec:errorbars}

Using the statistics of $g_{\eta_0}(\bm{y})$ has several advantages. One is that it reduces the impact of l.o.s. integration artifacts. 
In addition, it leads to proper error estimates for the empirical PDF. 

Introducing some function of $\eta_0$ noted $\varphi(\eta_0)$ which is expected to increase for increasing values of $|\eta_0|$,  the confidence interval above $(1-1/m^2)$ can be written as follows (see Bienayme-Tchebychev inequality, Eq. (\ref{eq:tchebyergodicestimate})):
\begin{equation}
 p(\eta_0) \equiv f_\eta(\eta_0)  = \hat{f}_L(\eta_0)  \left(1 \pm m \left(\varphi(\eta_0)\right)^{1/2} \left(\frac{l_c(\eta)}{R}\right)^{D/2} \right), \label{eq:errorbars}
\end{equation}
with $D=2$ and where
\begin{eqnarray}
\varphi(\eta_0) &=& \frac{\var{g_{\eta_0}}}{\hat{f}_L(\eta_0)^2 (\Delta \eta)^2 } \times \left(\frac{l_c(g_{\eta_0})}{l_c(\eta)}\right)^{2},  \\
&=&\frac{1}{\hat{f}_L(\eta_0) (\Delta \eta) } \times \left(\frac{l_c(g_{\eta_0})}{l_c(\eta)}\right)^{2} ,
\end{eqnarray}
because $\var{g_{\eta_0}} \simeq \hat{f}_L(\eta_0) (\Delta \eta)$ and where $l_c(g_{\eta_0})^2 \propto \Delta \eta$ so that the error bars on the PDF do not depend on the choice of bin size (for small  $\Delta \eta$, see App. \ref{app:PDF}). 

From the empirical ACF $C_{g_{\eta_0}}$, one can then estimate the correlation length of $g_{\eta_0}$ and thus $\varphi(\eta_0)$ for every $\eta_0$. 
Unfortunately, this procedure is hampered by the fact that the ACF becomes increasingly noisy at high contrasts $|\eta_0|>1$, due to sample sizes that are too small.

\begin{figure}[!t]
    \centering
    \includegraphics[trim=405 120 395 135, clip,width=\columnwidth]{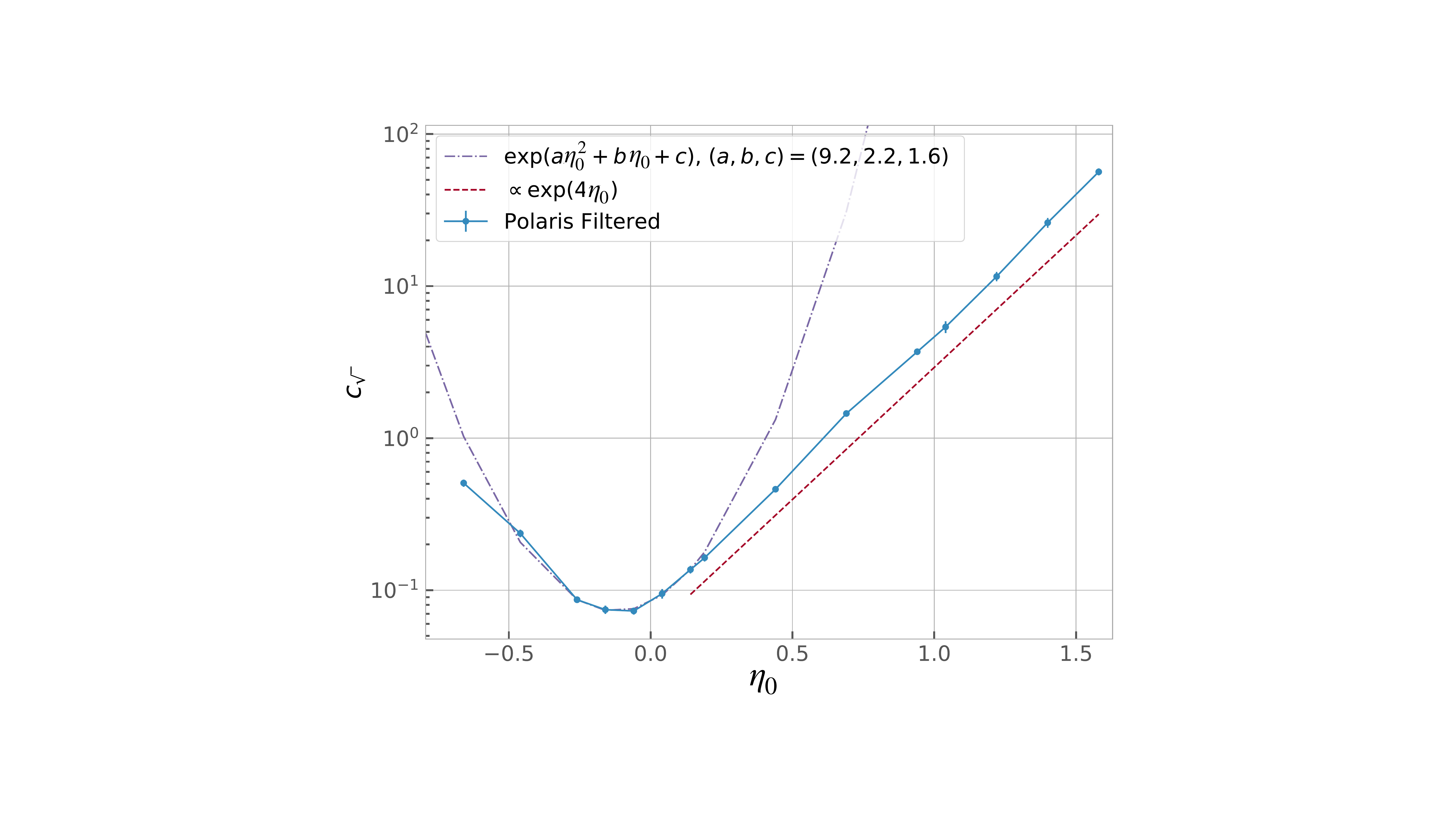}
    \caption{Constant of proportionality $c_{\sqrt{\,}}$ such that $C_{g_{\eta_0}}(\bm{y})=(f_\eta(\eta_0) \Delta \eta)^2 \times c_{\sqrt{\,}}/\sqrt{|\bm{y}|/L}$ at short scales. }
    \label{fig:constantprop}
\end{figure}

In principle, to determine $\varphi(\eta_0)$ and its variation one must calculate the complete integral that defines $l_c(g_{\eta_0})$. This may be avoided as follows.
The growth of $\varphi(\eta_0)$ may be obtained by   looking at the short scale behavior of the ACFs of $g_{\eta_0}$. 
In Fig.~\ref{fig:ACFofPDFestimators}, it appears that the values of $\hat{C}_{g_{\eta_0}}$ for positive column density contrasts ($\eta >0$) are isotropic and close to being $\propto |\bm{y}|^{-1/2}$ at short scales. 
We thus write that:
\begin{equation}
\hat{C}_{g_{\eta_0}}(\bm{y}) = (f_\eta(\eta_0) \Delta \eta)^2 \times c_{\sqrt{\,}} \, |\bm{y}/L|^{-1/2}
\end{equation}
where $c_{\sqrt{\,}}$ is a constant of proportionality that depends on $\eta_0$. Values of $c_{\sqrt{\,}}$ as a function of $\eta_0$ are given in Fig.~\ref{fig:constantprop}. 
We have only studied $g_{\eta_0}$ for $-0.7 \leq \eta_0 \leq 1.58$, because the ACFs are extremely noisy at high positive density contrasts ($\eta \geq 1.58$) due to poor sampling.  
At negative density contrasts ($\eta \leq -0.7$), where integration artifacts are the largest (see \S\ref{subsec:reducedeffects}), 
the ACFs are no longer sufficiently isotropic and do not conform to a scaling in $|\bm{y}|^{-1/2}$. 
Fig.~\ref{fig:constantprop} shows that $c_{\sqrt{\,}}(\eta_0)$ is an increasing function of $|\eta_0|$ for large $|\eta_0|$, illustrating the fact that 
$\varphi(\eta_0)$ is expected to be large compared to  $f_\eta(\eta_0)$ for large contrasts $|\eta_0|>1$.

\subsubsection{Correlation length of $g_{\eta_0}$ from Eq. (\ref{eq:estimateLcColumn})} \label{subsec:corrgInt}

 To obtain an estimate of $l_c(g_{\eta_0})$  and thus of $\varphi(\eta_0)$, we can use the results of Sec. \ref{sec:fluctuationsandcorr}. For each $\eta_0$ we compute the field $g_{\eta_0}(x,y)$ from the column density map $\eta(x,y)$. We then produce the two fields $\Sigma_{g_{\eta_0},x}(y)$ and $\Sigma_{g_{\eta_0},y}(x)$ obtained from the integration of $g_{\eta_0}(x,y)$ along the $x$ and $y$ direction, respectively:
\begin{eqnarray}
\Sigma_{g_{\eta_0},x}(y) = \int g_{\eta_0}(x,y) \mathrm{d} x, \\
\Sigma_{g_{\eta_0},y}(x) = \int g_{\eta_0}(x,y) \mathrm{d} y.
\end{eqnarray}
Computing the variance of these integrated fields we can obtain with  Eq. (\ref{eq:estimateLcColumn}) two estimates  $\hat{l}_{c,x}(g_{\eta_0})$ and $\hat{l}_{c,y}(g_{\eta_0})$ of $l_c(g_{\eta_0})$ within a factor of order unity:
\begin{eqnarray}
\frac{\hat{l}_{c,x}(g_{\eta_0})}{R_x} = \frac{\var{\Sigma_{g_{\eta_0},x}}}{\var{g_{\eta_0}}} \frac{1}{L_x^2}, \label{eq:estimg0x} \\
\frac{\hat{l}_{c,y}(g_{\eta_0})}{R_y} = \frac{\var{\Sigma_{g_{\eta_0},y}}}{\var{g_{\eta_0}}} \frac{1}{L_y^2}, \label{eq:estimg0y}
\end{eqnarray}
where $L_{x,y}=2 R_{x,y}$ are the lengths of the column density map in the $x$ and $y$ directions. For  the present map of Polaris these two lengths are approximately equal, $L_x \simeq L_y = L$. 

\begin{figure}[!t]
    \centering
    \includegraphics[width=\columnwidth]{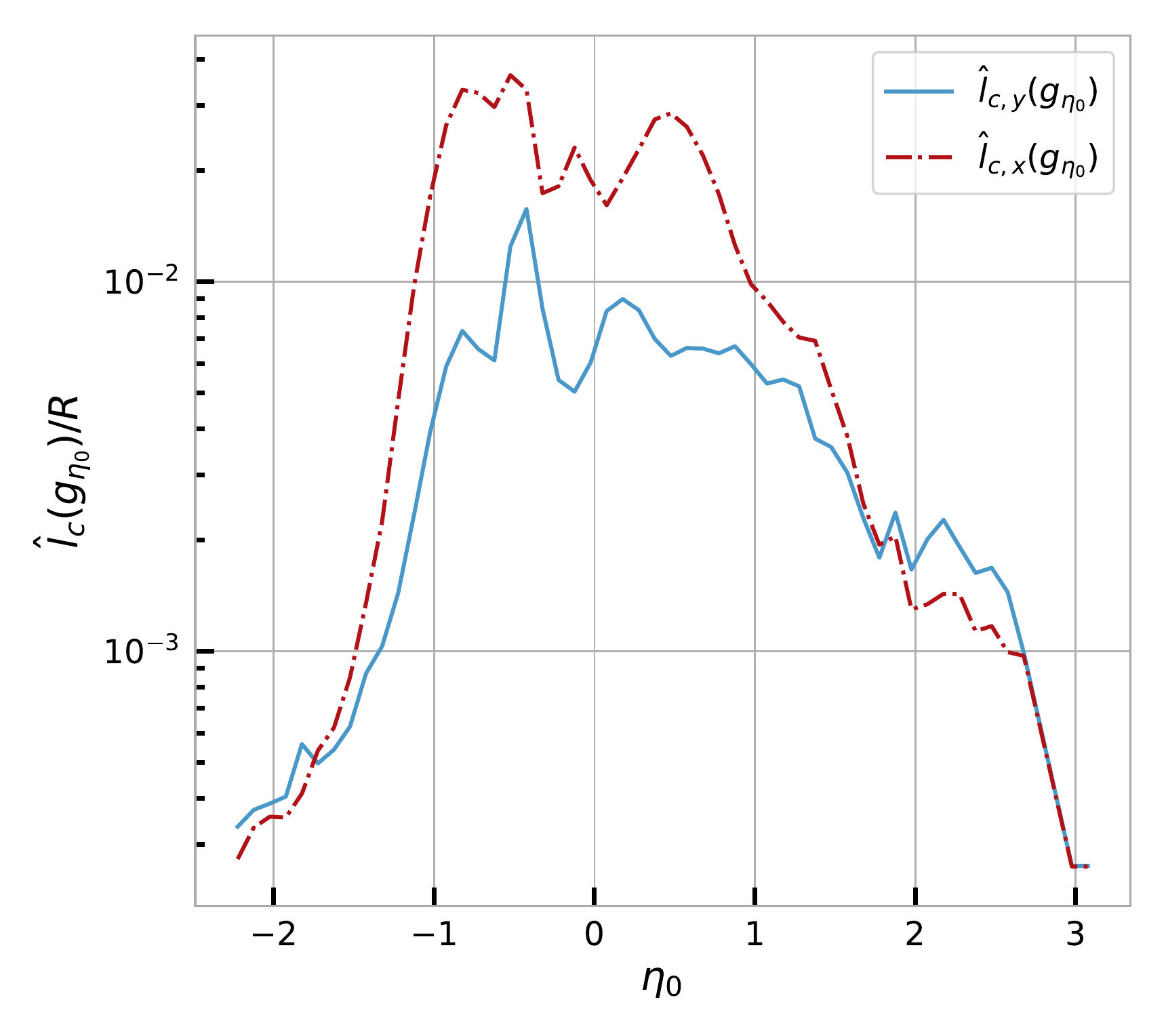}
    \caption{Estimate of the ratio  $l_c(g_{\eta_0})/R$ from Eqs. (\ref{eq:estimg0x}) and (\ref{eq:estimg0y}) at different $\eta_0$ for $\Delta \eta = 0.1$. The blue line gives the values  of estimate $\hat{l}_{c,y}(g_{\eta_0})$ while the red dash-dotted line gives the values of $\hat{l}_{c,x}(g_{\eta_0})$.  The two estimates always yield the same order of magnitude. }
    \label{fig:estimagelcg0}
\end{figure}

As, for $\eta_0>0$, the experimental ACFs of $g_{\eta_0}$ are fairly isotropic, we expect the above estimates of $l_c(g_{\eta_0})/R$ (Eqs. (\ref{eq:estimg0x}) and (\ref{eq:estimg0y})) to give similar and accurate results. They are given in Fig. \ref{fig:estimagelcg0}  for $\Delta \eta = 0.1$ and always yield the same order of magnitude.

\begin{figure}[!t]
    \centering
    \includegraphics[width=\columnwidth]{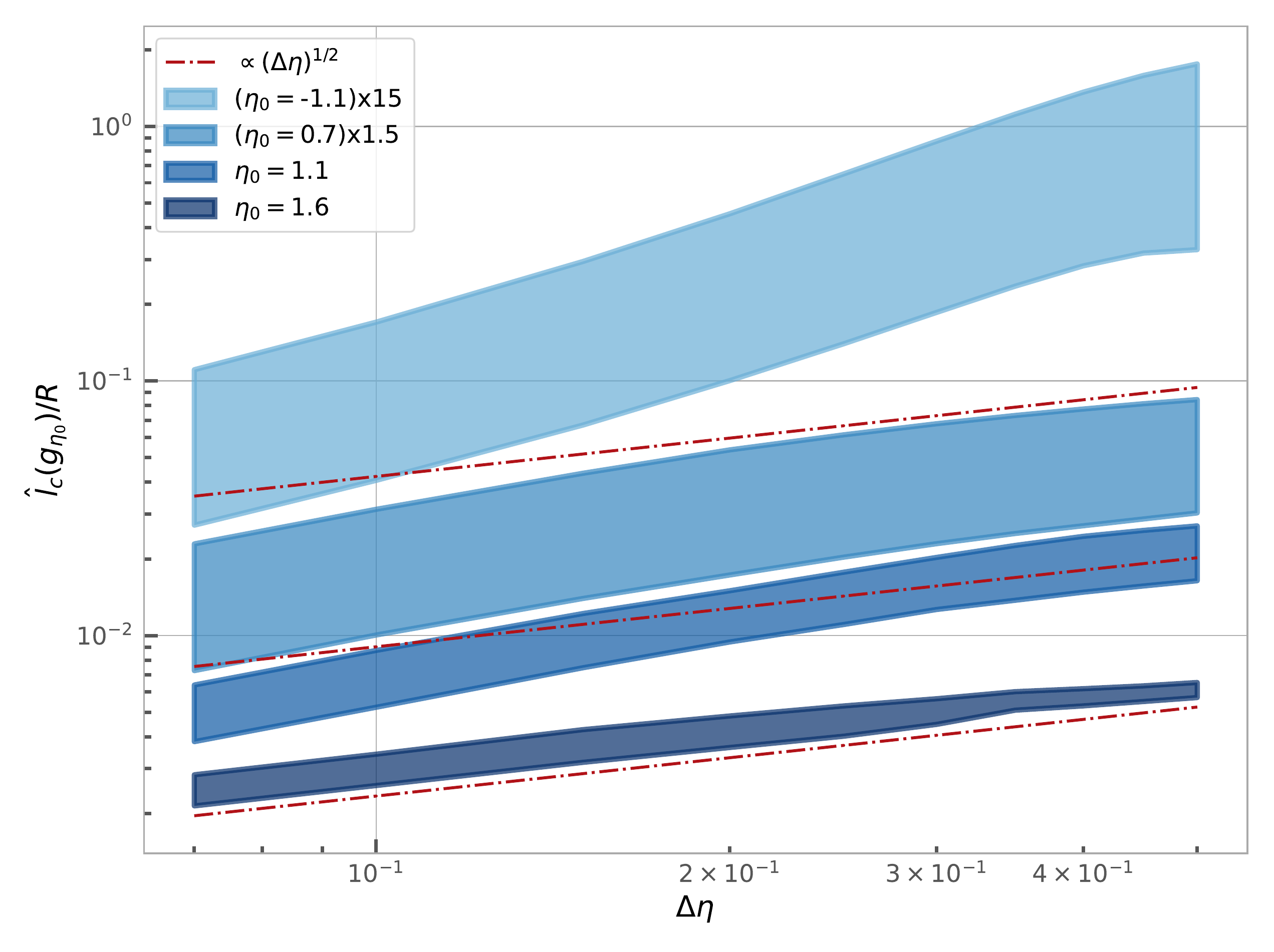}
    \caption{ Shadded regions bounded by the two estimates $\hat{l}_{c,x}(g_{\eta_0})$ and $\hat{l}_{c,y}(g_{\eta_0})$ for different bin sizes $\Delta \eta$ at four values of $\eta_0 = -1.1,$  $0.7,$ $1.1,$ $1.6$ from light to dark blue respectively. For clarity the values of the estimates for  $\eta_0 = -1.1$ and  $\eta_0 = 0.7$ were multiplied by $15$ and $1.5$ to shift their curves upward. Red dash-dotted lines indicate a scaling $\propto (\Delta \eta)^{1/2}$. At positive density contrasts $\eta >0$ the estimates follow a scaling close to the predicted $\hat{l}_{c,x,y}(g_{\eta_0}) \propto (\Delta \eta)^{1/2}$. }
    \label{fig:estimatelcg0Deltan}
\end{figure}

To test the accuracy of the estimates  $\hat{l}_{c,x}(g_{\eta_0})$ and $\hat{l}_{c,y}(g_{\eta_0})$ of $l_c(g_{\eta_0})$, we have tested how they scale with bin size $\Delta \eta$. If they were accurate estimates, they would have to conform to a scaling $\hat{l}_{c,x,y}(g_{\eta_0}) \propto (\Delta \eta)^{1/2}$ so that the error bars on the PDF (Eq. \ref{eq:errorbars}) do not depend on the choice of bin size (see App. \ref{app:PDF}). At positive density contrasts, $\eta >0$, the estimates $\hat{l}_{c,x}(g_{\eta_0})$ and $\hat{l}_{c,y}(g_{\eta_0})$ follow a scaling close to the predicted $\hat{l}_{c,x,y}(g_{\eta_0}) \propto (\Delta \eta)^{1/2}$. They do not, however, for negative contrasts, $\eta <0$, where the ACF is no longer isotropic and where there are strong integration artifacts. This can be seen on Fig. \ref{fig:estimatelcg0Deltan} where we display the shadded regions bounded by the two estimates $\hat{l}_{c,x}(g_{\eta_0})$ and $\hat{l}_{c,y}(g_{\eta_0})$ for different bin sizes $\Delta \eta$ at 4 values of $\eta_0 = -1.1,$  $0.7,$ $1.1,$ $1.6$.

We then conclude that  $\hat{l}_{c,x}(g_{\eta_0})$ and $\hat{l}_{c,y}(g_{\eta_0})$ can be used to accurately estimate $l_c(g_{\eta_0})$ for $\eta >0$, i.e. in the regions of interest for star formation. In practice we should assume that $l_c(g_{\eta_0})$ lies somewhere between $\hat{l}_{c,x}(g_{\eta_0})$ and $\hat{l}_{c,y}(g_{\eta_0})$ and compute error bars with both values. This is done in the next section.

\subsubsection{ Effective error bars on the observed PDF} \label{subsec:approxerrorbars}

 Once we have obtained these estimates of  $l_c(g_{\eta_0})/R$ and tested their accuracy, we can compute effective error-bars at a given confidence interval for the PDF $p(\eta_0) = f_\eta(\eta_0)$ with Bienayme-Tchebychev inequality, Eq. (\ref{eq:tchebyergodicestimate}):
\begin{equation}
p(\eta_0) \equiv f_\eta(\eta_0)  = \hat{f}_L(\eta_0)  \left(1 \pm m \frac{1}{\left(\hat{f}_L(\eta_0) \Delta \eta\right)^{1/2}} \frac{l_c(g_{\eta_0})}{R} \right), \label{eq:EstimatePDFerrorbars}
\end{equation}
with $\hat{f}_L(\eta_0)$ the estimate of the PDF produced by histograms of bin size $\Delta \eta$ and $m$ giving a confidence interval of over $1-1/m^2$.

\begin{figure}[!t]
    \centering
    \includegraphics[width=\columnwidth]{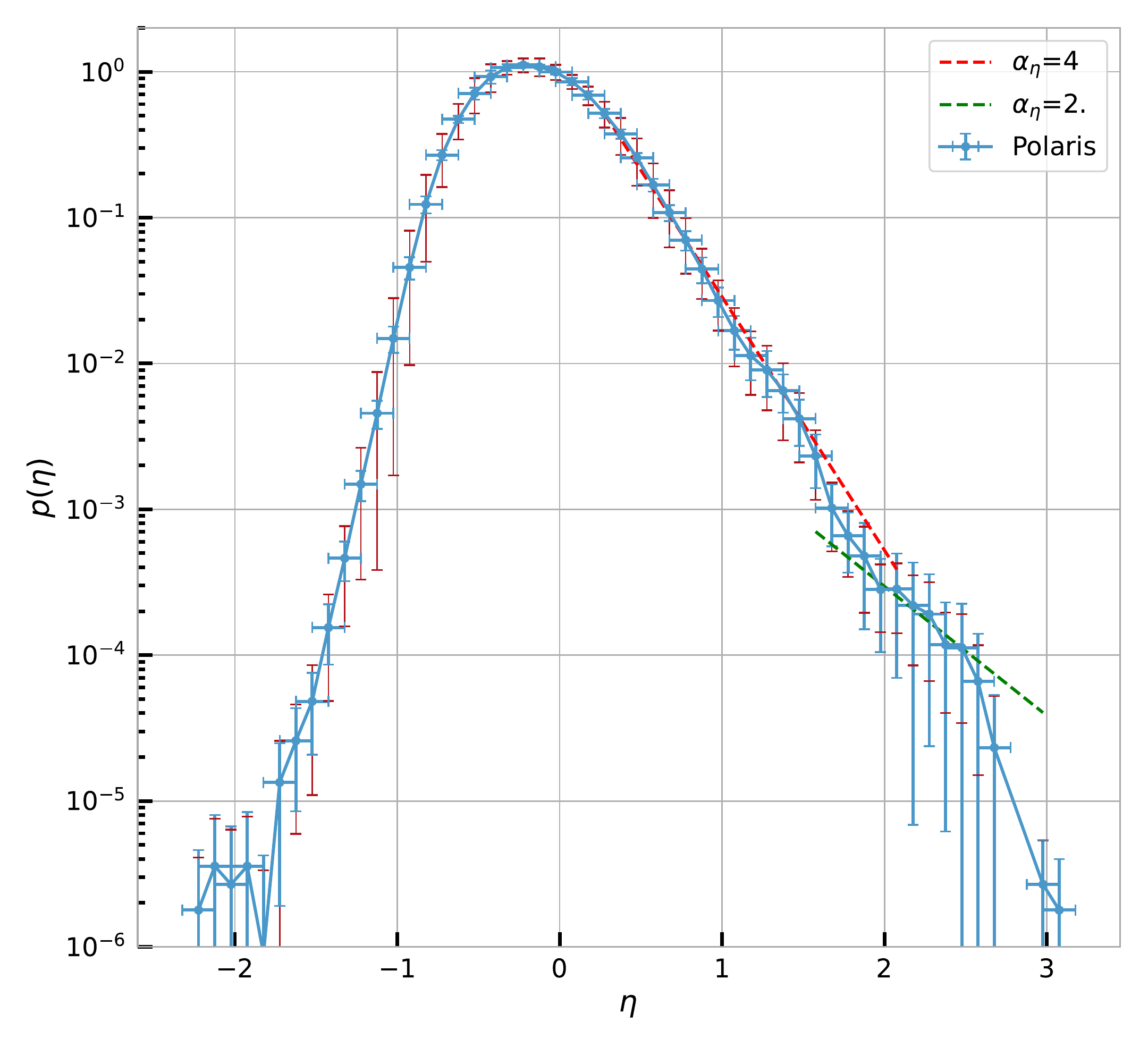}
    \caption{PDF of the logarithmic column-density $\eta$ with statistical error-bars for $m =2$ and the two estimates  $\hat{l}_{c,x}(g_{\eta_0})$ and $\hat{l}_{c,y}(g_{\eta_0})$ respectively in blue and red.    This emphasizes that  error bars should not be derived from Poisson statistics and that the accuracy of the low and high end parts of the PDF are degraded by the small sample size. }
    \label{fig:pdferrorbars}
\end{figure}

Fig.~(\ref{fig:pdferrorbars}) displays the empirical Polaris PDF, with error bars computed from Eq. (\ref{eq:EstimatePDFerrorbars}) for the two estimates  $\hat{l}_{c,x}(g_{\eta_0})$ and $\hat{l}_{c,y}(g_{\eta_0})$ with $\Delta \eta = 0.1$. We have taken $m=2$ to obtain a confidence interval of over $75\%$. As expected, the amplitudes of the error bars and thus of $\varphi(\eta_0)$ grow  with increasing values of $|\eta_0|$. These error bars may be inaccurate for $\eta_0<0$ because the estimates $\hat{l}_{c,x}(g_{\eta_0})$ and $\hat{l}_{c,y}(g_{\eta_0})$ show a dependence that is  too strong on bin size $\Delta \eta$ at these low column densities (see \S \ref{subsec:corrgInt}). However, they are accurate at high column densities $\eta_0>0$ and serve to emphasize that error bars should not be derived from Poisson statistics and that the accuracy of the low and high end parts of the PDF are severely degraded by sample sizes that are too small. 

\begin{figure*}[!t]
    \centering
    \includegraphics[trim=0 70 0 0, clip,width=\textwidth]{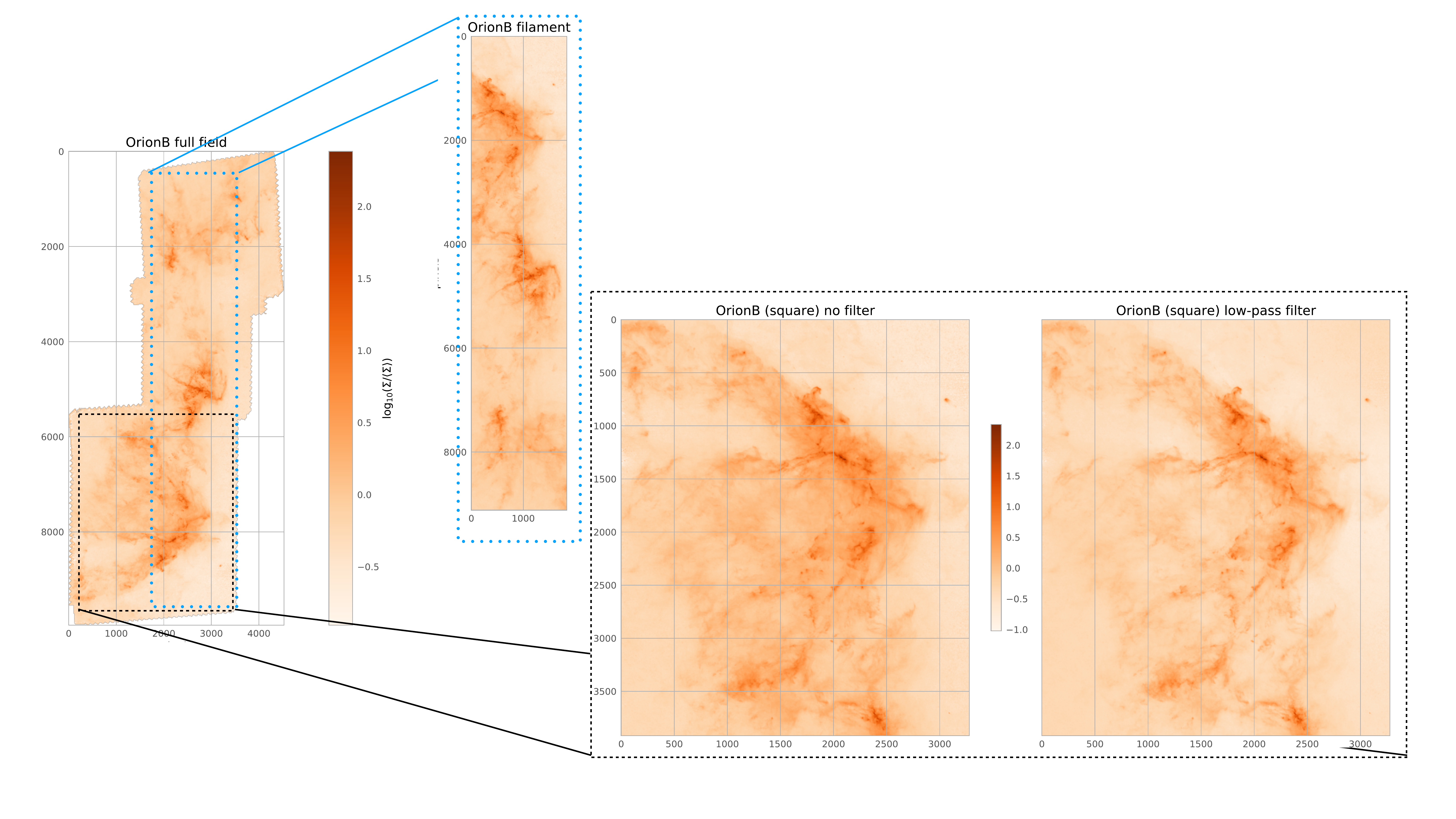}
    \caption{Column-density maps of the Polaris cloud. Left panel: Full observed field. Middle panel: Extracted filamentary region. Right panel: Extracted “square" region, unfiltered and low pass filtered (see \S \ref{sec:Obs}).}
    \label{fig:Orion_decoup}
\end{figure*}

\section{Applications to the Orion B cloud} \label{sec:OrionB}

In this section, we apply the results of  \S\ref{sec:Math} to the Orion B cloud \citep{schneider2013,orkisz2017}, another well studied star-forming MC. 
In this case, one encounters additional difficulties because the observed field is markedly elongated in the "vertical" direction ($y$) with data over a region whose geometrical shape 
is not suited to a straightforward data analysis (see Fig.~\ref{fig:Orion_decoup}). 
For this reason, we have extracted 2 parts of the cloud with rectangular shapes. One is elongated with a length that is close to the vertical dimension ($L_y$) of the total field of observation, which we shall refer to as a "filament". 
A second part is a rectangular one with an aspect ratio close to 1, with a length close to the maximum horizontal length of the full cloud ($L_x$), which we shall refer to as a "square" region (see Fig.~\ref{fig:Orion_decoup}). 
We determine the ACF of these two subregions and the associated correlation lengths in App.~\ref{app:OrionB} .

Using the ACF, we find that $l_c(\eta)/L_x \sim 10^{-1}$. When using the variances $\var{\rho}$ and $\var{\Sigma}$, we obtain a lower value: $l_c(\eta)/L_z \sim 10^{-2}$, 
where $L_z$ is the characteristic thickness of the cloud (along the line of sight).

\section{Conclusion}


In this article, we have examined the validity of statistical homogeneity and ergodicity when deriving general properties of star-forming molecular clouds from observations or numerical results of 
some of their properties. Notably, we have focused on the field of density fluctuations and its  PDF. This is a fundamental quantity since these fluctuations are believed to be at the root of the star formation process.
It is thus essential to examine the validity of a statistical approach in order to assess the accuracy of the  determination of the statistical properties of the cloud from the observations or simulations of a limited number  of samples. 
To fulfill  this goal, we first use the ergodic theory for any random field $X$  to derive some rigorous statistical results. 
We explain how to calculate the correlation length  of fluctuations in this field, $l_c(X)$, from the autocovariance function (ACF)  (Eq.~(\ref{eq:defLc})). 
We show that the estimation of the correlation length allows one to define an effective number of samples, $N$, such that a space (or time) average of a single realization is formally equivalent 
to averaging over $N$ independent samples (see e.g. \citealt{papoulis1965}).  When it is difficult to determine the correlation length from the empirical ACF, we have shown alternative ways to estimate it from fluctuations in Sec.~\ref{sec:fluctuationsandcorr}.  

We then apply this statistical approach and  the results of ergodic theory to astrophysical systems  in \S \ref{subsec:applicationtoastro}, \ref{sec:columndens} and \ref{sec:Obs}. In \S  \ref{subsec:applicationtoastro}, we examine in particular the stochastic fields induced by compressible turbulent motions driven at large scale. We show that while the energy of turbulent motion is injected at large scale, comparable to the whole system $L$, the correlation length of either the specific kinetic energy or the density is actually small compared to $L$ (see \S \ref{subsec:injdiffcorr}) . We stress that there is no contradiction in these results since the injection scale and correlation length are two completely different quantities.  

In \S  \ref{sec:columndens} we apply our results to the observed column-density field, which is related to the (volume) density field in the cloud. Applying the results of \S \ref{sec:fluctuationsandcorr},  we have devised a method to determine the correlation length, or more exactly the ratio of the correlation length over the size of the cloud (or the box of numerical simulations), from the variances of both the volume-density and column-density fields.  We have also shown that the statistics of the column-density field are affected by artifacts due to integration along the line of sight. These artifacts tend to generate an artificial anistropy in the colum density field and thus in the empirical ACF. Using the previous results, we have then examined in detail the Polaris cloud, which serves as a template for initial stages of star formation in MCs, in  \S \ref{sec:Obs}. We showed that the artificial anistropy in the empirical ACF results in an overestimation of the correlation length of density fluctuations within the cloud. Estimating the variance of the underlying density field, $\var{\rho/\left< \rho \right>}$, and computing the variance of the column-density field, $\var{\Sigma/\left< \Sigma \right>}$, we are able to derive a more accurate estimate of the correlation length $l_c$ (Eq.~(\ref{eq:relationratiovar})), which can be an order of magnitude smaller than the one obtained from the empirical ACF (\S\ref{subsec:ObservedACF}).

 Moreover, we have shown that studying the statistics of the PDF ergodic estimator for positive column-density contrasts enables us to get rid of most of the integration anisotropy bias (\S\ref{subsec:reducedeffects}).
 It also allows a proper evaluation of statistical error bars and shows that these  (i) cannot be derived from simple Poisson statistics and (ii) become increasingly large for increasing density contrasts ($|\eta|\geq 1$), severely reducing, in particular, the accuracy of the high end part of the PDF because of the small sample size (see Sec.~\ref{subsec:errorbars}). Furthermore, we provide a method that can be used by observers and numericists to determine robust error bars in Sec.~\ref{subsec:approxerrorbars}. 
 
 Finally, we found that the correlation length of the density field in the Polaris cloud is of about $\sim$1\% of the size of the cloud ($l_c(\rho)/R \sim 10^{-2}$). We have also examined the more complex Orion B cloud to confirm the results obtained for Polaris in \S \ref{sec:OrionB} .


These calculations provide a rigorous framework for the analysis of the global properties of star-forming clouds from limited statistical observations of their density and surface fluctuating properties.
They show in particular that  for typical star-forming clouds at the onset of the star formation process, the correlation length of density fluctuations is much smaller than the size of the cloud. This justifies the assumption and the relevance of a statistical approach based on statistical homogeneity when studying the PDF of the cloud (\citep{jaupart2020,jaupart2021generalized}), as done e.g. in cosmology or in the study of turbulence.

\begin{acknowledgements}
The authors are grateful to Christoph Federrath for always providing data from his numerical simulations upon request. The authors also thank Phillipe Andre for helpful discussions.
This research has made use of data from the Herschel Gould Belt survey (HGBS) project (http://gouldbelt-herschel.cea.fr). The HGBS is a Herschel Key Programme jointly carried out by SPIRE Specialist Astronomy Group 3 (SAG 3), scientists of several institutes in the PACS Consortium (CEA Saclay, INAF-IFSI Rome and INAF-Arcetri, KU Leuven, MPIA Heidelberg), and scientists of the Herschel Science Center (HSC).
\end{acknowledgements}

\bibliographystyle{aa} 
\bibliography{biblio.bib}

\begin{appendix}
\section{Ergodic estimate for a general control volume $\Omega$.} \label{app:ergodiccalculationgeneralvolume}
We described in Sec.~(\ref{sec:Math}) some known ergodic results, but they are derived for a cubic control volume $\Omega=[-\frac{L}{2},\frac{L}{2}]^D$. These results obviously do not depend on the shape of the control volume. We give here the general formulation for any control volume $\Omega$ possessing a center of symmetry (meaning that $\forall \bm{y} \in \Omega$, $-\bm{y} \in \Omega$). We again denote $|\Omega|$ the volume of $\Omega$ and define the linear size of $\Omega$ as $L^D=|\Omega|$. The ergodic estimate Eq.~(\ref{eq:defestimergodic}) is then
\begin{eqnarray}
\hat{X}_{\Omega} = \frac{1}{|\Omega|} \int_\Omega X(\bm{y}) \, \mathrm{d} \bm{y}.
\end{eqnarray}
To obtain its variance, one has to compute the double integral
\begin{eqnarray}
\var{\hat{X}_{\Omega}} &=& \frac{1}{|\Omega|^2} \iint_{\Omega^2} \esp{X(\bm{y})\, X(\bm{z}) - \esp{X}^2} \, \mathrm{d} \bm{y} \,\mathrm{d} \bm{z} \nonumber \\
&=& \frac{1}{|\Omega|^2} \iint_{\Omega^2} C_{X}(\bm{y}-\bm{z}) \, \mathrm{d} \bm{y} \,\mathrm{d} \bm{z}.
\end{eqnarray}
Using the change of variables $(\bm{u},\bm{v})=\varphi(\bm{y},\bm{z})=(\bm{y}-\bm{z},\bm{y}+\bm{z})$, one obtain
\begin{eqnarray}
\var{\hat{X}_{\Omega}} &=& \frac{1}{|\Omega|^2} \int_{2\,\Omega} C_X(\bm{u}) \int_{ \varphi_2^{\bm{u}}(\Omega)} \frac{\mathrm{d} \bm{u} \,\mathrm{d} \bm{v}}{2^D} 
\end{eqnarray}
where 
\begin{eqnarray}
\varphi_2^{\bm{u}}(\Omega) = 2 \left( (\Omega - \bm{u}) \cap \Omega \right) + \bm{u}
\end{eqnarray}
to obtain
\begin{eqnarray}
\var{\hat{X}_{\Omega}} &=& \frac{1}{|\Omega|} \int_{2\,\Omega} C_X(\bm{u}) \frac{|(\Omega - \bm{u}) \cap \Omega|}{|\Omega|} \, \mathrm{d} \bm{u}.
\end{eqnarray}
We then obtain the general Slutsky's theorem, $X$ is mean ergodic if and only if 
\begin{eqnarray}
\frac{1}{L^D} \int_{2\,\Omega} C_X(\bm{u})  \, \mathrm{d} \bm{u} \xrightarrow[L \rightarrow \infty]{} 0.
\end{eqnarray}

\section{Ergodic estimators of the CMF and PDF} \label{app:ergodicCMFPDF}
%
\subsection{Cumulative Distribution Function (CMF)} 
 

The CMF of the stochastic field $X$ can be constructed as the average of a particular function of the field $X$. Indeed, by definition, $F_X(x_0) = \mathbb{P}\left(X(\bm{y}) \leq x_0 \right)$ and a simple calculation shows that $\mathbb{P}\left(X(\bm{y}) \leq x_0 \right) = \esp{ h_{x_0}\left(X(\bm{y}) \right) }$ with $h_{x_0}(z) = \Theta(x_0 -z)$, where $\Theta$ is Heaviside step function. We are then ready to determine the confidence interval for the estimated CMF $F_X$ of $X$. To do so, we need to apply the results of Sec.~(\ref{sec:Math}) to the field $h_{x_0}(X(\bm{y}))$. The ‘‘natural" ergodic estimator of $F_X(x_0)$ is thus:
\begin{equation}
    \hat{F}_L(x_0) = \frac{1}{L^D} \int_{[-\frac{L}{2},\frac{L}{2}]^D} h_{x_0}\left(X(\bm{y}) \right) \, \mathrm{d} \bm{y}.
\end{equation}
Then, to obtain the variance of $\hat{F}_L(x_0)$ we need to express the ACF of $h_{x_0}\left(X(\bm{y}) \right)$. We have
\begin{equation}
    C_{h_{x_0}}(\bm{y}) = F_X^{(2)}(x_0,x_0,\bm{y})-F_X(x_0)^2
\end{equation}
where $F_X^{(2)}(x_0,x_0,\bm{y}) = \mathbb{P}\left(X(\bm{u}+\bm{y})) \leq x_0; \, \mathrm{and} \, X(\bm{u}) \leq x_0 \right) $ is the second-order distribution function and is the probability to have both $X(\bm{u}+\bm{y}) \leq x_0$ and $X(\bm{u}) \leq x_0 $. The variance of $\hat{F}_L(x_0)$ is then
\begin{eqnarray}
    \mathrm{Var}\left(\hat{F}_L(x_0)\right) &=& \frac{1}{(L)^D} \int_{[-L,L]^D} C_{h_{x_0}}(\bm{y}) \, \prod_{k=1}^D \left( 1 - \frac{|y_k|}{L} \right) \, \mathrm{d} \bm{y} \label{eq:varFL} \\
    &\simeq& C_{h_{x_0}}(\bm{0}) \left(\frac{l_c(h_{x_0})}{R}\right)^D \nonumber \\
    &=& F_X(x_0) \left(1 - F_X(x_0) \right) \left(\frac{l_c(h_{x_0})}{R}\right)^D, \label{eq:varCMF}
\end{eqnarray}
providing  $C_{h_{x_0}}$ is integrable so one can define $l_c(h_{x_0})$. Again, comparing with the result for a repeated trial experiment where $N$ samples of $X(\bm{y})$ are drawn (for the same point $\bm{y}$) shows that the ratio $(R/l_c(h_{x_0}))^D$ serves as an effective number $N$ of trials (see e.g. \citealt{papoulis1965}). 

For practical purpose and in order to give an interval of confidence, when $F_X$ is not known one can use the estimate $\hat{F}_L$ in Eq.~(\ref{eq:varCMF}) \citep{papoulis1965}. Furthermore, here, $l_c(h_{x_0})$ is a function of $x_0$ and cannot in general be simply estimated from $l_c(X)$. The length  $l_c(h_{x_0})$ can, however, be estimated by repeating the experiment several times and using the results of Sec.~(\ref{subsec:repeatedtrials}).
\newline

\subsection{Probability Density Function (PDF)} \label{app:PDF}

To build an estimator of the PDF $f_X(x_0)$ of $X$ we do not use the definition $f_X(x_0) = \esp{ \delta(X(\bm{y}) - x_0)}$ but the common approximation, suited for data analysis, $f_X(x_0) \Delta x \simeq F_X(x_0 + \Delta x) - F_X(x_0) = \esp{h_{x_0 + \Delta x}(X(\bm{y})) - h_{x_0}(X(\bm{y}))} $ for a sufficiently small bin spacing $\Delta x$. Noting $g_{x_0}(X(\bm{y}) = h_{x_0 + \Delta x}(X(\bm{y})) - h_{x_0}(X(\bm{y}))$ we build the estimator 
\begin{eqnarray}
\hat{f}_L(x_0) \Delta x = \frac{1}{L^D} \int_{[-\frac{L}{2},\frac{L}{2}]^D} g_{x_0}\left(X(\bm{y}) \right) \, \mathrm{d} \bm{y}.
\end{eqnarray}
The ACF of $g_{x_0}(X)$ is 
\begin{eqnarray}
    C_{g_{x_0}}(\bm{y}) &=& F_X^{(2)}(x_0 + \Delta x,x_0 + \Delta x,\bm{y}) + F_X^{(2)}(x_0 ,x_0 ,\bm{y}) \nonumber\\
    &&- F_X^{(2)}(x_0 + \Delta x,x_0,\bm{y}) - F_X^{(2)}(x_0 ,x_0 + \Delta x,\bm{y}) \nonumber \\
   && - \left(F_X(x_0 + \Delta x) - F_X(x_0) \right)^2 \label{eq:covgx0}
\end{eqnarray}
with
\begin{eqnarray}
    C_{g_{x_0}}(\bm{0}) &=& F_X(x_0 + \Delta x) - F_X(x_0) -  \left(F_X(x_0 + \Delta x) - F_X(x_0) \right)^2 \nonumber\\
    &\simeq& f_X(x_0) \Delta x \left( 1- f_X(x_0) \Delta x \right) + {\mathrm{O}}(\Delta x^{2}) \\
    &\simeq& f_X(x_0) \Delta x + {\mathrm{O}}(\Delta x^{2}).
\end{eqnarray}
We then know that a sufficient condition for $X$ to be density ergodic is either $C_{g_{x_0}}(\bm{y}) \xrightarrow[|\bm{y}| \rightarrow \infty]{ } 0$ or $C_{g_{x_0}}(\bm{y})$ is integrable. 

To find out how rapidly $C_{g_{x_0}}(\bm{y})$  decays to zero we note that
\begin{eqnarray}
 C_{g_{x_0}}(\bm{y}) &\simeq& \left( \frac{\partial^2 F_X^{(2)}}{\partial x_1 \partial x_2}(x_0,x_0,\bm{y})  - f_X(x_0)^2  \right) \Delta x^2 + {\mathrm{O}}(\Delta x^{3}) \label{eq:approxcovgx00} \\
   &=& \left(f_X^{(2)}(x_0,x_0,\bm{y}) - f_X(x_0)^2 \right) \Delta x^2 + {\mathrm{O}}(\Delta x^{3}), \label{eq:approxcovgx0}
\end{eqnarray}
where $f_X^{(2)}$ is the second-order density function. Eqs. \ref{eq:approxcovgx00} and \ref{eq:approxcovgx0} are only valid for $\bm{y} \neq \bm{0}$ because $f_X^{(2)}$ is degenerate for $\bm{y} = \bm{0}$ as $F_X^{(2)}(x_1,x_2,\bm{0}) = F_X(\mathrm{min}(x_1,x_2))$, where $\mathrm{min}(x_1,x_2)$ is not differentiable. The variance of the ergodic estimator $\hat{f}_{L,x_0}$ is, then,
\begin{eqnarray}
\var{\hat{f}_{L,x_0}} = \left(f_X(x_0) \Delta x \right) \left(\frac{l_c(g_{x_0})}{R}\right)^D,
\end{eqnarray}
where $l_c(g_{x_0})^D \propto \Delta x$ (see Eq.~(\ref{eq:approxcovgx0})).

\subsection{Gaussian process} \label{app:subsec:Gaussian}
If the field $X(\bm{y})$ is Gaussian we have 
\begin{equation}
    f_{X, G}^{(2)}(x_1,x_2, \bm{y})= \frac{1}{2 \pi \left| \underline{\bm{\Sigma}}(\bm{y})\right|^{1/2}} \exp \left( {- \frac{1}{2}(\bm{x}_\mu)^{T} \underline{\bm{\Sigma}}(\bm{y})^{-1} (\bm{x}_{\mu})} \right) 
\end{equation}
where $\bm{x}_\mu = (x_1 - \mu, x_2 - \mu)$, with $\mu = \esp{X }$, $\left| \underline{\bm{\Sigma}}(\bm{y})\right|$ is the determinant of the matrix $\underline{\bm{\Sigma}}(\bm{y})$  and
\begin{equation}
    \underline{\bm{\Sigma}}(\bm{y}) = \begin{pmatrix}
     \sigma(X)^2  & C_X(\bm{y}) \\
C_X(\bm{y}) &  \sigma(X)^2  
\end{pmatrix},
\end{equation}
We see that, as $\left| \underline{\bm{\Sigma}}(\bm{y})\right| = \sigma(X)^4 - C_X(\bm{y})^2 = \left(\sigma(X)^2 - C_X(\bm{y}) \right)\left(\sigma(X)^2 + C_X(\bm{y}) \right) $, $f^{(2)}$ is degenerate for $\bm{y}=\bm{0}$. However, for $\bm{y} \neq \bm{0}$, we have
\begin{eqnarray}
    f_{X, G}^{(2)}(x_0,x_0, \bm{y}) &=& \frac{1}{2 \pi \left| \underline{\bm{\Sigma}}(\bm{y})\right|^{1/2}} \exp \left( - x_{0,\mu}^2 \, \frac{\sigma(X)^2 - C_X(\bm{y})}{\sigma(X)^4 - C_X(\bm{y})^2} \right) \nonumber \\
    &=& \frac{1}{2 \pi \left| \underline{\bm{\Sigma}}(\bm{y})\right|^{1/2}} \exp \left( - \frac{\left(x_0 - \mu \right)^2}{\sigma(X)^2 + C_X(\bm{y})} \right)
\end{eqnarray}
Hence
\begin{small}
\begin{eqnarray}
C_{g_{x_0}}(\bm{y}) &\simeq&  \left(\frac{1}{\left(\left(1 + \frac{C_X(\bm{y})}{\sigma(X)^2}\right)\left(1 - \frac{C_X(\bm{y})}{\sigma(X)^2}\right)\right)^{1/2}} \exp \left(\frac{C_X(\bm{y}) (x_0 - \mu)^2}{ \sigma(X)^4\left(1 + \frac{C_X(\bm{y})}{\sigma(X)^2} \right)} \right) -1  \right) \nonumber \\   &&\times \frac{\Delta x^2}{2 \pi \sigma(X)^2} \exp \left(-\frac{(x_0-\mu)^2}{\sigma(X)^2}\right)  + {\mathrm{O}}(\Delta x^{3}).
 \end{eqnarray}
\end{small}
 Noting the normalized ACF $\Tilde{C}_X=C_X/C_X(\bm{0}) = C_X/\sigma(X)^2$ and the reduced variable $x_0^r = §(x_0-\mu)/\sigma(X)$ we have
\begin{eqnarray}
C_{g_{x_0}}(\bm{y}) &\simeq& \left(\frac{1}{\left(1-\Tilde{C}_X(\bm{y})^2\right)^{1/2}} \exp\left(\frac{\Tilde{C}_X(\bm{y}) (x_0^r)^2}{1 + \Tilde{C}_X(\bm{y})} \right) -1 \right) \nonumber \\
&&\times \frac{\Delta x^2}{2 \pi \sigma(X)^2} \exp \left(-(x_0^r)^2\right) + {\mathrm{O}}(\Delta x^{3}) \\
&\simeq& \left(\frac{1}{\left(1-\Tilde{C}_X(\bm{y})^2\right)^{1/2}} \exp\left(\frac{\Tilde{C}_X(\bm{y}) (x_0^r)^2}{1 + \Tilde{C}_X(\bm{y})} \right) -1 \right) \nonumber \\
&& \times f_X(x_0)^2 \, (\Delta x )^2 + {\mathrm{O}}(\Delta x^{3}). \label{eq:Cgx0ref}
\end{eqnarray}

\subsubsection{Integrability of the ACF and short scale analysis} \label{app:PDF_shortscale}
If $C_X$ decays to zero (as assumed) then for $|\bm{y}| \rightarrow \infty$, we have $C_{g_x0}(\bm{y}) \sim  \Tilde{C}_X(\bm{y}) \, (x_0^r)^2 \, f_X(x_0)^2 \, (\Delta x )^2 $. Thus, if $C_X$ is integrable, then so is $C_{g_x0}$ at $|\bm{y}| \rightarrow \infty$.

As mentioned above, Eq.~(\ref{eq:Cgx0ref}) is only valid for $|\bm{y}|>0$, so the divergence at $\bm{y}=\bm{0}$ is artificial as $C_{g_{x_0}}(\bm{0}) = f_X(x_0) \Delta x$ is finite. However, if Eq.~(\ref{eq:Cgx0ref}) is integrable at $\bm{y}=\bm{0}$, this ensures that the errors of approximation of $C_{g_{x_0}}$ near $\bm{y}=\bm{0}$ have a small effect on the estimation of $l_c(g_{x_0})$ (which is an integral). The divergence of Eq.~(\ref{eq:Cgx0ref}) at $\bm{y}=\bm{0}$ is given by
\begin{equation}
    \frac{1}{\left(1-\Tilde{C}_X(\bm{y})^2\right)^{1/2}} \exp \left(-\frac{1}{2}(x_0^r)^2\right) \times f_X(x_0)^2 \, (\Delta x )^2 \label{eq:shortscaledivergence}.
\end{equation}
For an exponential isotropic ACF this yields a divergence $\propto r^{-1/2}$, while for a differentiable field X with an ACF being isotropic at short scales this yields a divergence $\propto r^{-1}$. Thus, in most cases for $D\geq 2$ Eq.~(\ref{eq:Cgx0ref}) is integrable at $|\bm{y}| \rightarrow 0$.

Computing the integral of $C_{g_{x_0}}(\bm{y})$ is not straightforward for any decaying and integrable ACF $C_X(\bm{y})$. Expanding the  exponential in Eq.~(\ref{eq:Cgx0ref}), we have 
\begin{eqnarray}
     \exp\left(\frac{\Tilde{C}_X(\bm{y}) (x_0^r)^2}{1 + \Tilde{C}_X(\bm{y})} \right) &=& 1+ \sum_{n\geq 1} \frac{(x_0^r)^{2n}}{n!} \left(\frac{\Tilde{C}_X(\bm{y})}{1 + \Tilde{C}_X(\bm{y})} \right)^n . 
\end{eqnarray}
We then have to specify or bound the integrals
\begin{eqnarray}
\frac{1}{2^D} \int_{\mathbb{R}^D} \left(\frac{\Tilde{C}_X(\bm{y})}{1 + \Tilde{C}_X(\bm{y})} \right)^n \, \frac{\mathrm{d} \bm{y}}{\left(1-\Tilde{C}_X(\bm{y})^2\right)^{1/2}} &=& l_c(X)^D \, c_n, \\
 \frac{1}{2^D}  \int_{\mathbb{R}^D} \left(\frac{1}{\left(1-\Tilde{C}_X(\bm{y})^2\right)^{1/2}} -1 \right) \mathrm{d} \bm{y} &=& l_c(X)^D \, c_0,
\end{eqnarray}
to obtain 
\begin{equation}
  \frac{1}{2^D}  \int_{\mathbb{R}^D} C_{g_{x_0}}(\bm{y}) \,\mathrm{d} \bm{y} = l_c(X)^D \varphi(x_0^r) \times  f_X(x_0)^2 \, (\Delta x )^2 + {\mathrm{O}}(\Delta x^{3}),
\end{equation}
where $\varphi(x_0^r)$ is a function of $x_0$ which we need to bound to obtain a confidence interval. A lower bound of $\varphi(x_0^r)$ can be obtained due to the convexity of the exponential:
\begin{eqnarray}
 \varphi(x_0^r) \geq c_0 + c_1 (x_0^r)^2
\end{eqnarray}
For general monotonic decreasing  (hence positive) ACFs, the study of the functions $\frac{x}{1+x} \frac{1}{(1-x^2)^{1/2}}$ and $\frac{1}{(1-x^2)^{1/2}}-1$ shows that $c_0 \gtrsim 0.1$ and $c_1 \geq 0.77$.
\subsubsection{Exponential ACF}
To go a little further and obtain a formula that will help to grasp some expected features of the ergodic estimate of $g_{x_0}$ we study the special case of an exponential ACF. For the present study we limit ourselves to the case $D\leq2$. Then if $\Tilde{C}_X$ is an (isotropic) exponential, $\Tilde{C}_X(y) = \exp(-|y|/\lambda)$, we can bound the integral of Eq.~(\ref{eq:Cgx0ref}).  Indeed,  for $n\geq 1$,
\begin{eqnarray}
\frac{l_c(X)^D}{2^{n-1}} \geq  \frac{1}{2^D} \int_{\mathbb{R}^D} \left(\frac{\Tilde{C}_X(\bm{y})}{1 + \Tilde{C}_X(y)} \right)^n \times \frac{\mathrm{d} \bm{y}}{\left(1-\Tilde{C}_X(\bm{y})^2\right)^{1/2}}.
\end{eqnarray}
We then have
\begin{eqnarray}
 \frac{1}{2^D} \int_{\mathbb{R}^D} C_{g_{x_0}}(\bm{y}) \, \mathrm{d} \bm{y} &\leq& l_c(X)^D \left(2 \exp\left(\frac{(x_0^r)^2}{2}\right) + \tilde{c}_0^D -2 \right) \nonumber \\ &&\times \left(f_X(x_0) \Delta x \right)^2 + {\mathrm{O}}(\Delta x^{3}) 
\end{eqnarray}
where $\tilde{c}_0^D= \mathrm{ln}(2)$ and $0.17$ for $D=1$ and $D=2$, respectively. This gives an upper bound to the correlation length of $g_{x_0}$ but  overestimates its value for $|x_0^r| \gg 1$. However, near the average ($|x_0^r| \ll 1 $) we can  approximate
\begin{eqnarray}
   \frac{1}{2^D} \int_{\mathbb{R}^D} C_{g_{x_0}}(\bm{y}) \, \mathrm{d} \bm{y} &\simeq& l_c(X)^D \left(\tilde{c}_1^D \, (x_0^r)^2 + \tilde{c}_0^D \right) \nonumber \\
    &&\times \left(f_X(x_0) \Delta x \right)^2 + {\mathrm{O}}(\Delta x^{3}), \label{eq:applcapproxnearav}
\end{eqnarray}
where  $\tilde{c}_1^D=1$ and $0.88$ for $D=1$ and $D=2$, respectively. We note that, due to the convexity of the exponential, the right hand side of Eq.~(\ref{eq:applcapproxnearav}) is actually a lower bound of the integral $\forall x_0$.


We can then construct a confidence interval with more than $1-1/m^2$ of confidence such that the true $f_X(x_0) $ lies in
\begin{equation}
  f_X(x_0)  = \hat{f}_L(x_0)  \left(1 \pm m \left(\varphi(x_0^r)\right)^{1/2} \left(\frac{l_c(X)}{R}\right)^{D/2} \right), \nonumber
\end{equation} 
where
\begin{equation}
  \tilde{c}_1^D \, (x_0^r)^2 + \tilde{c}_0^D \leq \varphi(x_0^r) \leq 2 \exp\left(\frac{(x_0^r)^2}{2}\right) + \tilde{c}_0^D -2. 
\end{equation} 
Using the lower bound to approximate $\varphi(x_0^r)$, $\varphi(x_0^r) \simeq \tilde{c}_1^D \, (x_0^r)^2 + \tilde{c}_0^D$, while accurate for $|x_0^r|\ll 1$, is most probably an underestimation for $|x_0^r|\gg 1$.  However, it allows to show that the statistics of events that deviate largely from the mean needs an increasingly large sample size to have a high degree of confidence.

\subsection{Deterministic function of a Gaussian field.} \label{app:determinsticfunction}
The results derived in Sec.~(\ref{app:subsec:Gaussian}) can be extended to the case where $X(\bm{y}) = \psi(S(\bm{y}))$ with $\psi$ a diffeomorphism  and $S$ a Gaussian field. A particular example is that of a lognormal field where $\psi = \exp$. We call this function $\psi$ a deterministic function because statistical properties of the field $X$ can be obtained from those of $S$. Indeed, for such a field $X$, the first and second order distribution functions read
\begin{eqnarray}
    f_X(x_0) = \frac{f_S(s_0)}{|(\psi^{-1})'(x_0)|} \label{eq:deterministicdensitypdf} \\
    f_X^{(2)}(x_1,x_2; \bm{y}) = \frac{f_S^{(2)}(s_1,s_2;\bm{y})}{|(\psi^{-1})'(x_1)| |(\psi^{-1})'(x_2)|},
\end{eqnarray}
where $s_j= \psi^{-1}(x_j)$ \citep{papoulis1965}. Without loss of generality we can further assume that the field $S$ is centered with  variance unity. We note that the function $\psi$ can be obtained by inverting Eq.~(\ref{eq:deterministicdensitypdf}). Indeed, if only $f_X$ and $f_S$ are known we can obtain $\psi$ by realizing that $\psi^{-1}$ verifies the differential equation:
\begin{equation}
   |(\psi^{-1})'(x_0)| = \frac{f_S( (\psi^{-1})(x_0))}{f_X(x_0)}.
\end{equation}
If one further assumes that $\psi$ is an increasing diffeomorphism $(\psi^{-1})' \geq 0$, one obtains
\begin{equation}
    s_0=\psi^{-1}(x_0)=\sqrt{2} \, \mathrm{erf}^{-1} \left( -1 + 2 \int_{x_{\rm min}}^{x_0} f_X(x) \, \mathrm{d} x \right),
\end{equation}
where $x_{\rm min}$ is the minimum value that can be taken by the field $X$ and $\mathrm{erf}^{-1}$ is the inverse of the error function. The use of this equation requires a high precision on $f_X$ due to the large variation of $\mathrm{erf}^{-1}$, which is complicated in general.

Then the ACF of $X$ can be obtained by performing the integral:
\begin{eqnarray}
C_X(\bm{y}) = \int \psi(s_1)\psi(s_2) \left(f_S^{(2)}(s_1,s_2;\bm{y}) - f_S(s_1)\, f_S(s_2) \right) \, \mathrm{d}s_1\, \mathrm{d}s_2. \nonumber
\end{eqnarray}
Then Eq.~(\ref{eq:Cgx0ref}) becomes
\begin{eqnarray}
C_{g_{x_0}}(\bm{y}) &\simeq& \left(\frac{1}{\left(1-\Tilde{C}_S(\bm{y})^2\right)^{1/2}} \exp\left(\frac{\Tilde{C}_S(\bm{y}) (s_0)^2}{1 + \Tilde{C}_S(\bm{y})} \right) -1 \right) \nonumber \\
&& \times f_X(x_0)^2 \, (\Delta x )^2 + {\mathrm{O}}(\Delta x^{3}).
\end{eqnarray}

\subsubsection{Log-normal fields}
For a log-normal field $\rho = \exp(s)$, $\psi = \exp$, $\psi^{-1}=\mathrm{ln}$ and  $s$ is not centered ($\esp{s} \neq 0$) and does not have a variance unity ($\sigma(s) \neq 1$), in general. Then the calculation of the ACF yields:
\begin{eqnarray}
C_\rho(\bm{y})= \esp{\rho}^2 \left( e^{C_s(\bm{y})} -1 \right).
\end{eqnarray}
As a consequence, because $e^{ax} -1 \leq x(e^{a}-1)$ for $0\leq x\leq 1$ and $ax \leq e^{ax} -1$ $\forall x$, if $C_\rho$ (or $C_s$) is monotonically decaying to $0$ then 
\begin{equation}
    \left(\frac{\sigma(s)^2}{e^{\sigma(s)^2}-1}\right)^{1/3} l_c(s) \leq l_c(\rho) \leq l_c(s).
\end{equation}
In typical star forming conditions $\sigma(s)^2 \lesssim 4$, giving
\begin{equation}
    0.4 \, l_c(s) \lesssim l_c(\rho) \leq l_c(s),
\end{equation}
or
\begin{equation}
    l_c(s) \sim l_c(\rho).
\end{equation}
This suggests that as long as $\var{X} = \var{\psi(S)}$ is not too big, one can expect to have $l_c(X) \sim l_c(S)$.

\section{Orion B cloud} \label{app:OrionB}


\subsection{ACF of the square and filament region}
\begin{figure*}
    \centering
    \includegraphics[trim=0 70 0 100, clip,width=\textwidth]{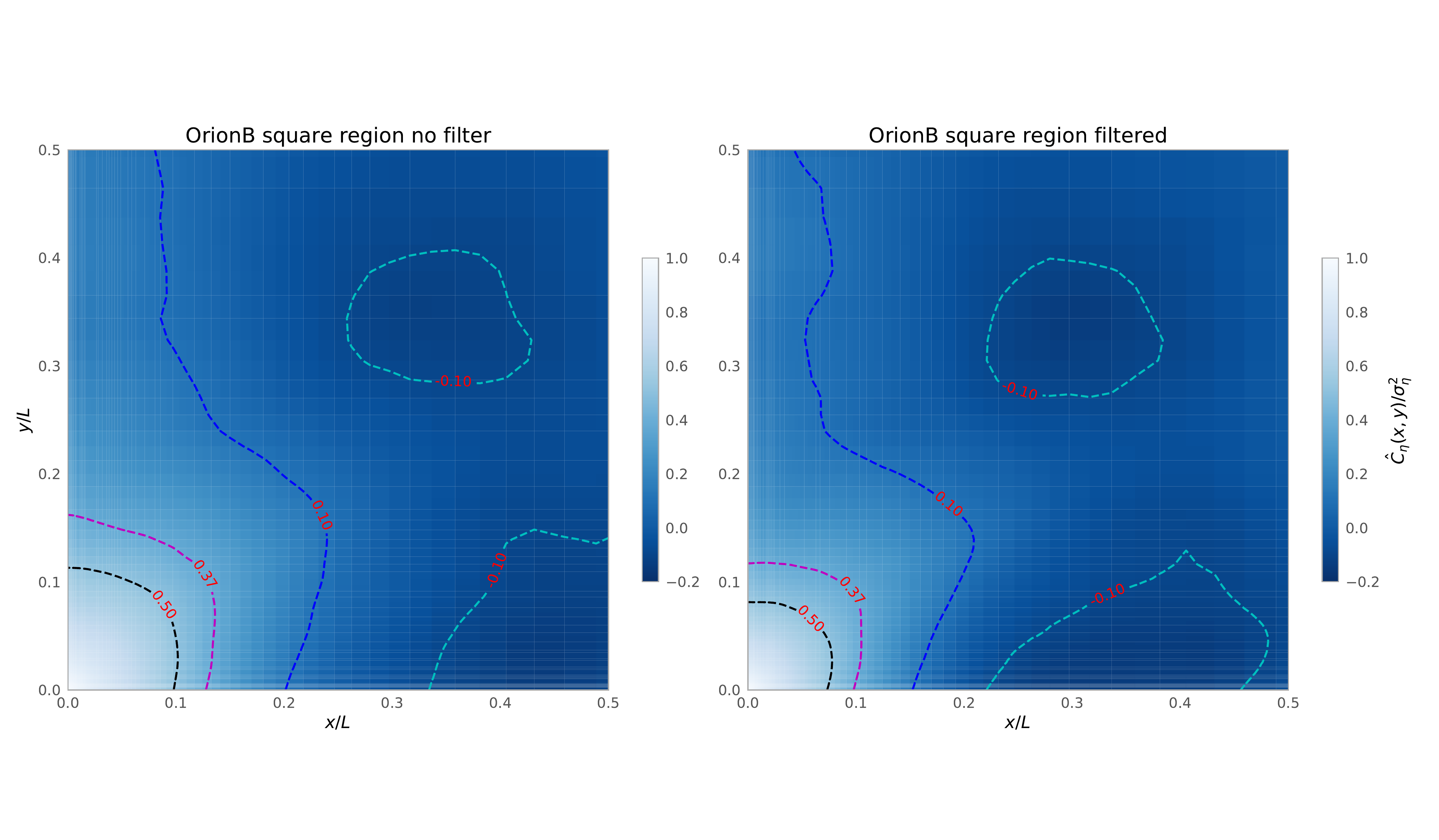}
    \caption{ACF of the “square region". Left panel: unfiltered.  Right panel: low pass filtered up to scale $L/2$. Again, filtering large-scale gradients reduces the anisotropy at short scales and reduces the estimated correlation length.}
    \label{fig:Orion_squareACF}
\end{figure*}

\begin{figure}
    \centering
    \includegraphics[angle =270, trim=750 0 110 0, clip, width=\columnwidth]{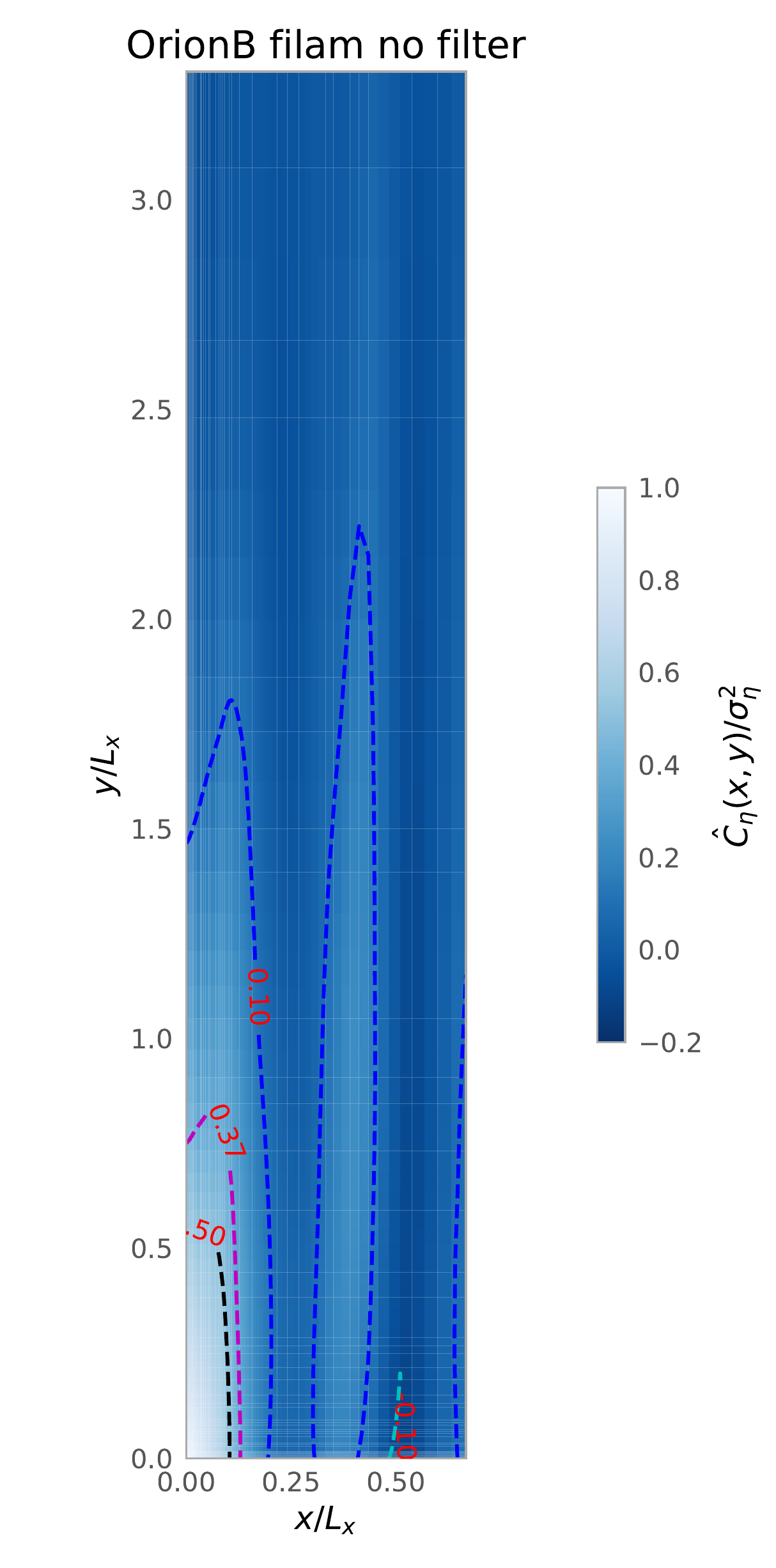}
    \caption{ACF of the unfiltered “filament region". A strong anisotropy is present in the $y$-direction.}
    \label{fig:Orion_filamACF}
\end{figure}

We computed the ACF of the unfiltered and  low pass filtered square region (up to scale $L/2$, see \S \ref{sec:Obs}), as well as the (unfiltered) “filament" region. The results are presented in Figs.~(\ref{fig:Orion_squareACF}) and (\ref{fig:Orion_filamACF}). Filtering large-scale gradients reduces again the anisotropy at short scales and reduces the estimated correlation length.

\begin{figure}
    \centering
    \includegraphics[width=\columnwidth]{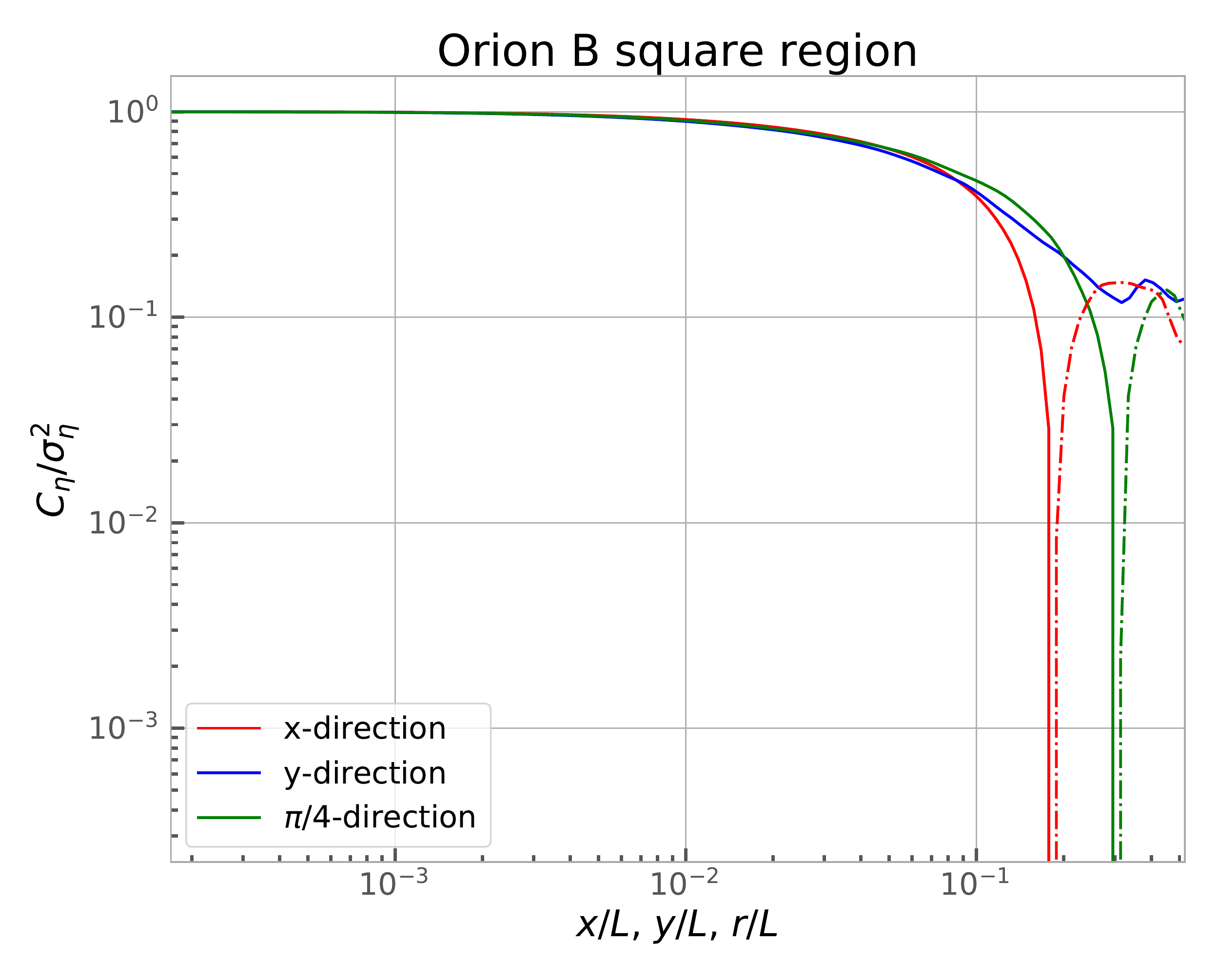}
    \caption{Reduced ACF of the low pass filtered map in three different directions. Red line: $x$-direction ($y=0$). Blue line: $y$-direction $(x=0)$. Green line: $\pi/4$ or $x=y$-direction. Dash dotted lines represent the values of the ACF when it is negative. A strong anisotropy is present in the $y$ direction at large scales ($y/L \geq   10^{-1}$).}
    \label{fig:Orion_ACF3DIR}
\end{figure}

To have a closer look at the behavior of the ACF of Orion B, we display in Fig.~(\ref{fig:Orion_ACF3DIR})  the reduced ACF of the low pass filtered map in 3 different directions, $x$ ($\theta=0$), $x=y$ ($\theta=\pi/4$) and $y$ ($\theta=\pi/2$). As seen from the heat maps but also from fig.~(\ref{fig:Orion_ACF3DIR}),  a strong anisotropy is present and located in the $y$ direction at large scales ($y/L \geq  \, 10^{-1}$). The resulting estimated correlation length $\hat{l}_c(\eta)$ is  of the order $l_c(\eta)/L_x \simeq  10^{-1}$.

\subsection{Correlation length from the variance of the column densities.}

As seen from Sec.~(\ref{subsec:variancecolumndensity}) and Eq.~(\ref{eq:relationratiovar}), one can also give an estimate of the ratio $l_c(\rho)/R$ by (1) computing the variance $\var{\Sigma/\esp{\Sigma}}$, (2) giving an estimate of $\var{\rho/\esp{\rho}}$ and (3) giving an estimate of the average thickness of the cloud (along the line of sight) $L_z$.  Here Orion B appears as a very elongated structure, and we will therefore only assume that $L_y \geq L_z \gtrsim L_x$ (with $L_y \simeq 3-4 \, L_x$). 

From observations of column densities we obtain $\var{\Sigma/\left<\Sigma\right>} \simeq 1.1$ while the PDF of column densities, exhibiting a power-law tail of exponent $\alpha_\eta=-2$ \citep{schneider2013,jaupart2020}, indicates an underlying density PDF with a power-law tail of exponent $-3/2$ implying a large variance. As For Polaris, running the power-law tail from $s=8$ to $s=s_{\rm ad} \simeq 16$ yields a variance $\var{\rho/\esp{\rho}}= 40$ and $\var{\rho/\esp{\rho}} \simeq 2300$ respectively (see \citealt{jaupart2020}). This yields a ratio $l_c(\rho) / L_z \lesssim10^{-2}$.

\end{appendix}

\end{document}